\documentclass[epj]{svjour}
\usepackage{amsmath,amssymb}
\usepackage{times}
\usepackage{txfonts}
\usepackage{dcolumn}
\usepackage{bm}
\usepackage{lineno}
\usepackage{graphicx}

\newcommand{\natu}{$^\mathrm{nat}$}

\newcommand{\etal}{\textit{et al}}
\newcommand{\ie}{\textit{i.e.}}
\newcommand{\eg}{\textit{e.g.}}
\usepackage{lscape}

\usepackage{longtable}
\DeclareSymbolFont{EulerExtension}{U}{euex}{m}{n}
\DeclareMathSymbol{\euintop}{\mathop} {EulerExtension}{"52}
\DeclareMathSymbol{\euointop}{\mathop} {EulerExtension}{"48}

\everymath{\displaystyle}

\begin{document}
\title{
Energy dependence of isomeric ratios for alpha-particle-induced reactions on natural platinum up to 50~MeV studied by simultaneous decay curve analysis
}

\author{
Naohiko Otuka\inst{1,2}\thanks{n.otsuka@iaea.org}
\and
Masayuki Aikawa\inst{2,3,4,5}\thanks{aikawa@sci.hokudai.ac.jp}
\and
S\'{a}ndor Tak\'{a}cs\inst{6}\thanks{stakacs@atomki.hu}
\and
Damdinsuren Gantumur\inst{7}
\and
Shuichiro Ebata\inst{8}
\and
Lkhagvasuren Bold\inst{7}
\and
Akihiro Nambu\inst{2}
\and
Hiromitsu Haba\inst{2}
}
\institute{
Nuclear Data Section,
Division of Physical and Chemical Sciences,
Department of Nuclear Sciences and Applications,
International Atomic Energy Agency,
A-1400 Wien, Austria
\and
Nishina Center for Accelerator-Based Science,
RIKEN,
Wako 351-0198,
Japan
\and
Faculty of Science,
Hokkaido University,
Sapporo 060-0810,
Japan
\and
Graduate School of Biomedical Science and Engineering,
Hokkaido University,
Sapporo 060-8638,
Japan
\and
Global Center for Biomedical Science and Engineering,
Faculty of Medicine,
Hokkaido University,
Sapporo 060-8648,
Japan
\and
HUN-REN Institute for Nuclear Research (ATOMKI),
H-4026 Debrecen,
Hungary
\and
School of Engineering and Technology,
National University of Mongolia,
Ulaanbaatar 14200,
Mongolia
\and
Graduate School of Science and Engineering,
Saitama University,
Saitama 338-8570,
Japan
}
\date{Received: date / Revised version: date}

\abstract{
The isomeric ratios and cross sections were measured up to 50~MeV for production of $^{198m,g}$Au, $^{197m,g}$Hg, $^{195m,g}$Hg and other Hg, Au and Pt radionuclides in irradiation of natural platinum by alpha-particles using the stacked-foil activation technique and offline gamma-ray spectrometry.
The simultaneous decay curve analysis was applied to the emission rates of more than 40 gamma lines to directly derive the independent cross sections of all products without determining the cumulative cross sections.
The measured cross sections and isomeric ratios were compared with those compiled in the EXFOR library and simulated by TALYS-2.0.
We found that the \natu Pt($\alpha$,x)$^{198g+m}$Au, $^{197g+m}$Hg and $^{195g+m}$Hg cross sections are fairly reproduced by the simulation,
while the simulation result requires adjustment of the spin cutoff parameter for better reproduction of the isomeric ratios. 
}

\maketitle
\onecolumn
\section{Introduction}
The production of metastable states of some nuclides with cyclotrons has drawn attention because of their medical applications,
stimulating the measurement and evaluation of their excitation functions.
Production of $^{99m}$Tc and $^{44m}$Sc for SPECT application~\cite{Tarkanyi2019a,Hermanne2023a} and $^{34m}$Cl, $^{52m}$Mn, $^{82m}$Rb, $^{94m}$Tc and $^{110m}$In productions for PET application~\cite{Tarkanyi2019b,Hermanne2023b} are such examples.
Performance of nuclear reaction model codes for isomer productions is important in the optimization of target composition and irradiation conditions to maximize the yield of the radionuclide of interest and also to minimize the undesired presence of co-produced radionuclides.
%
%
To validate prediction of isomer productions by reaction model codes,
high quality experimental isomer production cross sections must be available, preferably in the EXFOR library~\cite{Otuka2014}.

When the half-life of the ground state is longer than the half-life of the metastable state and the difference in their half-lives is large,
we can assume that the metastable state had completely decayed to the ground state when the ground state activity was measured after long cooling,
and this simplifies determination of the isomeric ratio.
If the half-lives of the metastable and ground states are close to each other,
or the half-life of the metastable state is longer,
experimental determination of the isomeric ratio becomes less trivial.
Overestimation of the ground state production cross section due to insufficient difference in the half-lives of an isomer pair is discussed in the context of so-called ``supracumulative" cross section~\cite{Titarenko2002}.
We performed systematic compilation of the experimental isomeric ratios and noticed that the EXFOR library shows large deviation between measurements in excitation functions of some isomer production reactions~\cite{Rodrigo2023}.

Isomer production in the nuclear reaction codes has been modelled with spin dependence of the level density characterized by the spin cutoff parameter~\cite{Huizenga1960}.
Comparison of the experimental isomeric ratios with those simulated by TALYS-2.0~\cite{Koning2023} indicates that the best performance of the simulation is globally achieved when we reduce the spin cutoff parameter to half of the rigid body value~\cite{Rodrigo2023},
and this trend has been confirmed by subsequent works~\cite{Cannarozzo2023,Jacob2025}.

We have recently developed a tool performing the simultaneous decay curve analysis~\cite{Otuka2024},
which provides us the independent ground state production cross section directly without additional correction due to feeding by the metastable state or other decaying radionuclides.
We applied this procedure to determination of the \natu Pt($\alpha$,x)$^{198m,g}$Au, $^{197m,g}$Hg and $^{195m,g}$Hg isomer production cross sections~\cite{Otuka2024},
and found that this approach works well.
We also found that the measured isomeric ratios prefer reduction of the spin cutoff parameter in general.
However,
the maximum irradiation energy in our previous measurement (29~MeV) was too low to discuss performance of the reaction model in the energy region where the $^{198m,g}$Au and $^{195m,g}$Hg production cross sections exhibit their peaks.

The purpose of the present work is to study the performance of the model prediction for the isomer production cross sections determined with Pt foils irradiated by an alpha-particle beam at higher energies.

\section{Experimental}
Metallic foils of \natu Pt (21.672~mg/cm$^2$, nominal purity of 99.95\%), \natu Ti (2.343 mg/cm$^2$, nominal purity of 99.6\%) and $^{27}$Al (1.817 mg/cm$^2$, nominal purity of $>$99\%) were cut into 8$\times$8~mm$^2$ to form a target stack.
The Ti foils were interleaved to examine the beam flux and energy loss of the beam particles in the stack by using the \natu Ti($\alpha$,x)$^{51}$Cr reaction as the monitor reaction, while the Al foils were interleaved to catch the nuclei produced in the adjacent Pt foil that recoiled.
The molar masses recommended by IUPAC~\cite{Meija2016} and adopted by us are 195.084~g/mol for Pt and 47.867~g/mol for Ti.
In total, 19 Pt foils, 19 Al foils and 38 Ti foils were arranged in a stack of 19 sets of Pt-Al-Ti-Ti.
They were stacked in a target holder, which also served as a Faraday cup.
The beam energies at the upstream and downstream side of each foil were calculated by using the SRIM code~\cite{Ziegler2010} and their mid energy was adopted as the representative energy of the foil.
 
The target stack was irradiated for 1802~sec by an alpha-particle beam collimated to 3~mm in diameter and extracted from the RIKEN AVF cyclotron.
The initial beam energy of 50.8$\pm$0.2~MeV was determined by the time-of-flight method~\cite{Watanabe2014},
where the uncertainty was propagated from the uncertainties in the flight time and flight path length.
The intensity of the beam current was monitored by recording the charge integrated in the Faraday cup periodically and the stability of the beam flux during irradiation was confirmed with the average beam intensity of 6.3606$\times$10$^{11}$ alphas/sec ($\sim$204~nA).
 
Measurement of the activities of the irradiated foils was started 52~min after the end of irradiation using a high-purity germanium detector (ORTEC GEM30P4-70) connected to an ORTEC signal processing unit,
and the spectra were analysed by two software tools (SEIKO EG\&G Gamma Studio, Canberra Genie 2000).
The geometry-dependent detection efficiency was determined experimentally at various gamma energies by using a multiple point-like calibration source consisting of $^{57,60}$Co, $^{85}$Sr, $^{88}$Y, $^{109}$Cd, $^{113}$Sn $^{137}$Cs, $^{139}$Ce, $^{203}$Hg and $^{241}$Am at each distance between the detector surface and sample position used for the foil activity measurements.
The functional form of the efficiency curve at each distance was assumed to be
\begin{equation}
\ln \epsilon = \sum_{i=0}^5 a_i \left(\ln E\right)^i
\end{equation}
and the coefficients $a_i$ ($i$=0,5) were adjusted to reproduce the measured detection efficiencies at the calibration energies.
The measured peak area was corrected for the absorption of the gamma-ray in the foil assuming the fraction of the gamma-rays not absorbed in the foil is expressed by
\begin{equation}
f=\frac{1-e^{-\mu L}}{\mu L},
\end{equation}
where $\mu$ is the mass attenuation coefficient recommended by NIST~\cite{Hubbell2004}.
For the gamma lines adopted by us,
$f$ ranges between 0.945 (for $E_\gamma$=98~keV) and 0.999 (for $E_\gamma$=1468~keV).

The Pt and Al foil pairs and the second Ti monitor foil of each set were measured seven times or less.
In the present work, production cross sections were derived from the measured net peak areas, detection efficiencies, irradiation and target parameters as well as the decay data summarized in Table~\ref{tab:decaydata}.
These decay data were extracted from the ENSDF library~\cite{A200,A199,A198,A197,A196,A195,A194,A193,A192,A51} by the IAEA LiveChart of Nuclides~\cite{Verpelli2011} except for the decay branching ratios and gamma emission probabilities of $^{197m}$Hg and $^{195m}$Hg, for which we adopted the result of our previous work~\cite{Otuka2024}.
Considering the time scale of our measurement, an excitation level with a half-life shorter than 1 min is not treated as a metastable state in this work.
The gamma lines italicized in Table~\ref{tab:decaydata} were initially selected for the analysis but finally excluded as discussed later.

\begin{longtable}{llllllll}
\caption{
Half-life $T_{1/2}$, decay branching ratio $p$, gamma energy $E_\gamma$ and gamma emission probability $I$ adopted in the present work taken from the ENSDF library~\cite{A200,A199,A198,A197,A196,A195,A194,A193,A192,A51} except for the $^{197m}$Hg and $^{195m}$Hg decay branching ratios and gamma emission probabilities determined by us~\cite{Otuka2024}.
``Route" gives the production route via the reaction having the lowest threshold energy as well as decay included in our modelling.
All gamma lines listed here and not italicized were used in determination of the cross sections and isomeric ratios shown in the figures and table.
The gamma energy in the ENSDF library is rounded to one digit after the decimal point.
The threshold energy $E_\mathrm{thr}$ was calculated by TCalc~\cite{Shimizu2022,Shimizu2024} with AME2020 mass evaluation~\cite{Huang2021}.
The parenthesized number is the uncertainty corresponding to the last digits of the estimate, \eg, 72.4(24) means 72.4$\pm$2.4.
}
\label{tab:decaydata}
\\
\hline
Nuclide     & $T_{1/2}$      & $p$ (\%)             & $E_\gamma$ (keV) & $I$ (\%)           & $E_\mathrm{thr}$ (MeV) & Route                           & Ref.        \\
\hline                                                                                                                                                
$^{200m}$Au & 18.7 (5) h     & IT 16(1)             & \textit{133.2}   & \textit{0.52 (10)} & 14.7                   & $^{198}$Pt($\alpha$,d)          &~\cite{A200} \\
            &                &                      & 332.8            & 2.2 (5)            &                        &                                 &             \\
\cline{3-5}                                                                                
            &                & $\beta^-$ 84(1)      & 255.9            & 72.4 (24)          &                        &                                 &             \\
            &                &                      & 367.9            & 79.3 (9)           &                        &                                 &             \\
            &                &                      & \textit{497.8}   & \textit{81.8 (10)} &                        &                                 &             \\
            &                &                      & \textit{579.3}   & \textit{82.4 (10)} &                        &                                 &             \\
            &                &                      & 759.5            & 74.6 (24)          &                        &                                 &             \\
\cline{1-7}                                                                                
$^{200g}$Au & 48.4 (3) min   & $\beta^-$ 100        & 367.9            & 19 (3)             & 13.6                   & $^{198}$Pt($\alpha$,d)          &             \\
            &                &                      &                  &                    &                        & $^{200m}$Au (IT)                &             \\
\hline                                                                                     
$^{199m}$Hg & 42.67 (9) min  & IT 100               & 158.3            & 52.3 (10)          & 1.4                    & $^{195}$Pt($\alpha$,$\gamma$)   &~\cite{A199} \\
            &                &                      & 374.1            & 13.8 (11)          &                        &                                 &             \\
\cline{1-7}                                                                                
$^{199}$Au  & 3.139 (7) d    & $\beta^-$ 100        & 158.4            & 40.0 (7)           & 8.6                    & $^{196}$Pt($\alpha$,p)          &             \\
            &                &                      & 208.2            & 8.72 (18)          &                        &                                 &             \\
\hline                                                                                     
$^{198m}$Au & 2.272 (16) d   & IT 100               & 180.3            & 49 (5)             & 9.1                    & $^{195}$Pt($\alpha$,p)          &~\cite{A198} \\
            &                &                      & 204.1            & 39 (5)             &                        &                                 &             \\
            &                &                      & 214.9            & 77.0 (10)          &                        &                                 &             \\
            &                &                      & 333.8            & 18 (4)             &                        &                                 &             \\
\cline{1-7}                                                                                
$^{198g}$Au & 2.6941 (2) d   & $\beta^-$ 100        & 411.8            & 95.62 (6)          & 8.2                    & $^{195}$Pt($\alpha$,p)          &             \\
            &                &                      &                  &                    &                        & $^{198m}$Au (IT)                &             \\
\hline                                                                                     
$^{197m}$Hg & 23.8 (1) h     & IT 94.5(7)           & \textit{134.0}   & \textit{34.6 (3)}  & 10.4                   & $^{194}$Pt($\alpha$,n)          &~\cite{Otuka2024,A197} \\
\cline{3-5}                                                                                
            &                & EC 5.5(7)            & 279.0            & 3.9 (7)            &                        &                                 &             \\
\hline
$^{197g}$Hg & 64.14 (5) h    & EC 100               & 191.4            & 0.632 (22)         & 10.1                   & $^{194}$Pt($\alpha$,n)          &~\cite{A197} \\
            &                &                      &                  &                    &                        & $^{197m}$Hg (IT)                &             \\
\cline{1-7}                                                                                
$^{197m}$Pt & 95.41 (18) min & IT 96.7(4)           & 346.5            & 11.1 (3)           & 8.1                    & $^{198}$Pt($\alpha$,n+$\alpha$) &             \\
\cline{3-5}                                                                                
            &                & EC 3.3(4)            & 279              & 2.4 (6)            &                        &                                 &             \\
\cline{1-7}                                                                                
$^{197g}$Pt & 19.8915(19) h  & $\beta^-$ 100        & 191.4            & 3.7 (4)            & 7.7                    & $^{198}$Pt($\alpha$,n+$\alpha$) &             \\
            &                &                      &                  &                    &                        & $^{197m}$Pt (IT)                &             \\
\hline                                                                                     
$^{196m}$Au & 9.6 (1) h      & IT 100               & 147.8            & 43.5 (15)          & 15.1                   & $^{195}$Pt($\alpha$,t)          &~\cite{A196} \\
            &                &                      & 188.3            & 30.0 (15)          &                        &                                 &             \\
            &                &                      & 316.2            & 3.0 (3)            &                        &                                 &             \\
\cline{1-7}                                                                                
$^{196g}$Au & 6.1669(6) d    & EC/$\beta^+$ 93.0(3) & 333.0            & 22.9 (9)           & 14.5                   & $^{195}$Pt($\alpha$,t)          &             \\
            &                &                      & 355.7            & 87 (3)             &                        & $^{196m}$Au (IT)                &             \\
\hline                                                                                     
$^{195m}$Hg & 41.6(8) h      & IT 48.9(18)          &                  &                    & 11.3                   & $^{192}$Pt($\alpha$,n)          &~\cite{Otuka2024,A195} \\
\cline{3-5}                                                                                
            &                & EC/$\beta^+$51.1(18) & 207.1            & 0.41 (9)           &                        &                                 &             \\
            &                &                      & \textit{261.8}   & \textit{35 (4)}    &                        &                                 &             \\
            &                &                      & 279.3            & 0.16 (5)           &                        &                                 &             \\
            &                &                      & 368.6            & 0.38 (4)           &                        &                                 &             \\
            &                &                      & 386.4+           & 2.75 (25)          &                        &                                 &             \\
            &                &                      & 387.9            &                    &                        &                                 &             \\
            &                &                      & 560.3            & 7.9 (7)            &                        &                                 &             \\
\hline
$^{195g}$Hg & 10.53(3) h     & EC/$\beta^+$ 100     & 180.1            & 1.95 (24)          & 11.1                   & $^{192}$Pt($\alpha$,n)          &~\cite{A195} \\
            &                &                      & 207.1            & 1.6 (3)            &                        & $^{195m}$Hg (IT)                &             \\
            &                &                      & \textit{261.8}   & \textit{1.6 (3)}   &                        &                                 &             \\
            &                &                      & 585.1            & 2.0 (2)            &                        &                                 &             \\
            &                &                      & 599.7            & 1.83 (22)          &                        &                                 &             \\
            &                &                      & 779.8            & 7.0 (8)            &                        &                                 &             \\
            &                &                      & \textit{930.9}   & \textit{0.43(6)}   &                        &                                 &             \\
            &                &                      & 1111.0           & 1.48 (22)          &                        &                                 &             \\
\cline{1-7}                                                                                
$^{195}$Au  & 186.01(6) d    & EC 100               & 98.9             & 11.21 (15)         & 8.8                    & $^{192}$Pt($\alpha$,p)          &             \\
            &                &                      &                  &                    &                        & $^{195m}$Hg (EC/$\beta^+$)      &             \\
            &                &                      &                  &                    &                        & $^{195g}$Hg (EC/$\beta^+$)      &             \\
\cline{1-7}                                                                                
$^{195m}$Pt & 4.010(5) d     & IT 100               & 98.9             & 11.7 (8)           & 0.3                    & $^{195}$Pt($\alpha$,$\alpha'$)  &             \\
\hline                                                                                     
$^{194}$Hg  & 447(52) y      & EC 100               &                  &                    & 0                      & $^{190}$Pt($\alpha$,$\gamma$)   &~\cite{A194} \\
\cline{1-7}                                                                                
$^{194}$Au  & 38.02(10) h    & EC/$\beta^+$ 100     & 293.5            & 10.9 (3)           & 15.1                   & $^{192}$Pt($\alpha$,d)          &             \\
            &                &                      & 328.5            & 62.8 (16)          &                        & $^{194}$Hg (EC)                 &             \\
            &                &                      &                  &                    &                        &                                 &             \\
            &                &                      & 1468.9           & 6.80 (20)          &                        &                                 &             \\
\hline                                                                                                                                                              
$^{193m}$Hg & 11.8 (2) h     & IT 7.2(5)            &                  &                    & 12.3                   & $^{190}$Pt($\alpha$,n)          &~\cite{A193} \\
\cline{3-5}                                                                                
            &                & EC/$\beta^+$ 92.8(5) & \textit{258.0}   &                    &                        &                                 &             \\
            &                &                      & \textit{381.6}   &                    &                        &                                 &             \\
            &                &                      & \textit{499.7}   &                    &                        &                                 &             \\
            &                &                      & \textit{573.3}   &                    &                        &                                 &             \\
            &                &                      & \textit{932.4}   &                    &                        &                                 &             \\
\cline{1-7}                                                                                
$^{193g}$Hg & 3.80 (15) h    & EC/$\beta^+$ 100     & \textit{186.6}   & \textit{15.0 (20)} & 12.1                   & $^{190}$Pt($\alpha$,n)          &             \\
            &                &                      & \textit{258.0}   & \textit{8.7 (20)}  &                        & $^{193m}$Hg (IT)                &             \\
            &                &                      & \textit{381.6}   & \textit{15 (5)}    &                        &                                 &             \\
            &                &                      & \textit{581.0}   & \textit{4.0 (21)}  &                        &                                 &             \\
            &                &                      & 861.1            & 12.1 (24)          &                        &                                 &             \\
            &                &                      & 1118.8           & 7.8 (14)           &                        &                                 &             \\
\cline{1-7}                                                                                
$^{193}$Au  & 17.65(15) h    & EC/$\beta^+$ 100     & \textit{186.2}+  & \textit{10.57}     & 9.0                    & $^{193m}$Hg (EC/$\beta^+$)      &             \\
            &                &                      & \textit{187.8}   &                    &                        & $^{193g}$Hg (EC/$\beta^+$)      &             \\
            &                &                      & 255.6            & 6.5                &                        &                                 &             \\
\hline                                                                                                                                                              
$^{192}$Hg  & 4.85(20) h     & EC 100               & 99.4             & 0.68 (18)          & 19.4                   & $^{190}$Pt($\alpha$,2n)         &~\cite{A192} \\
            &                &                      & 157.2            & 7.2 (4)            &                        &                                 &             \\
            &                &                      & \textit{186.4}   & \textit{3.4 (6)}   &                        &                                 &             \\
            &                &                      & 204.6            & 0.85 (18)          &                        &                                 &             \\
            &                &                      & \textit{262.6}   & \textit{0.68 (22)} &                        &                                 &             \\
            &                &                      & 274.8            & 52 (4)             &                        &                                 &             \\
            &                &                      & 279.2            & 0.43 (6)           &                        &                                 &             \\
\cline{1-7}                                                                                
$^{192}$Au  & 4.94(9) h      & EC/$\beta^+$ 100     & 296.0            & 23(3)              & 15.6                   & $^{190}$Pt($\alpha$,d)          &             \\
            &                &                      & 316.5            & 59(7)              &                        & $^{192}$Hg (EC)                 &             \\
            &                &                      & 759.1            & 1.68(20)           &                        &                                 &             \\
\hline                                                                                     
$^{51}$Cr   & 27.704(3) d    & EC 100               & 320.1            & 9.91 (1)           & 0                      & $^{47}$Ti($\alpha$,$\gamma$)    &~\cite{A51}  \\
\hline
\end{longtable}

\begin{figure}[hbtp]
\begin{center}
\includegraphics[width=0.8\textwidth]{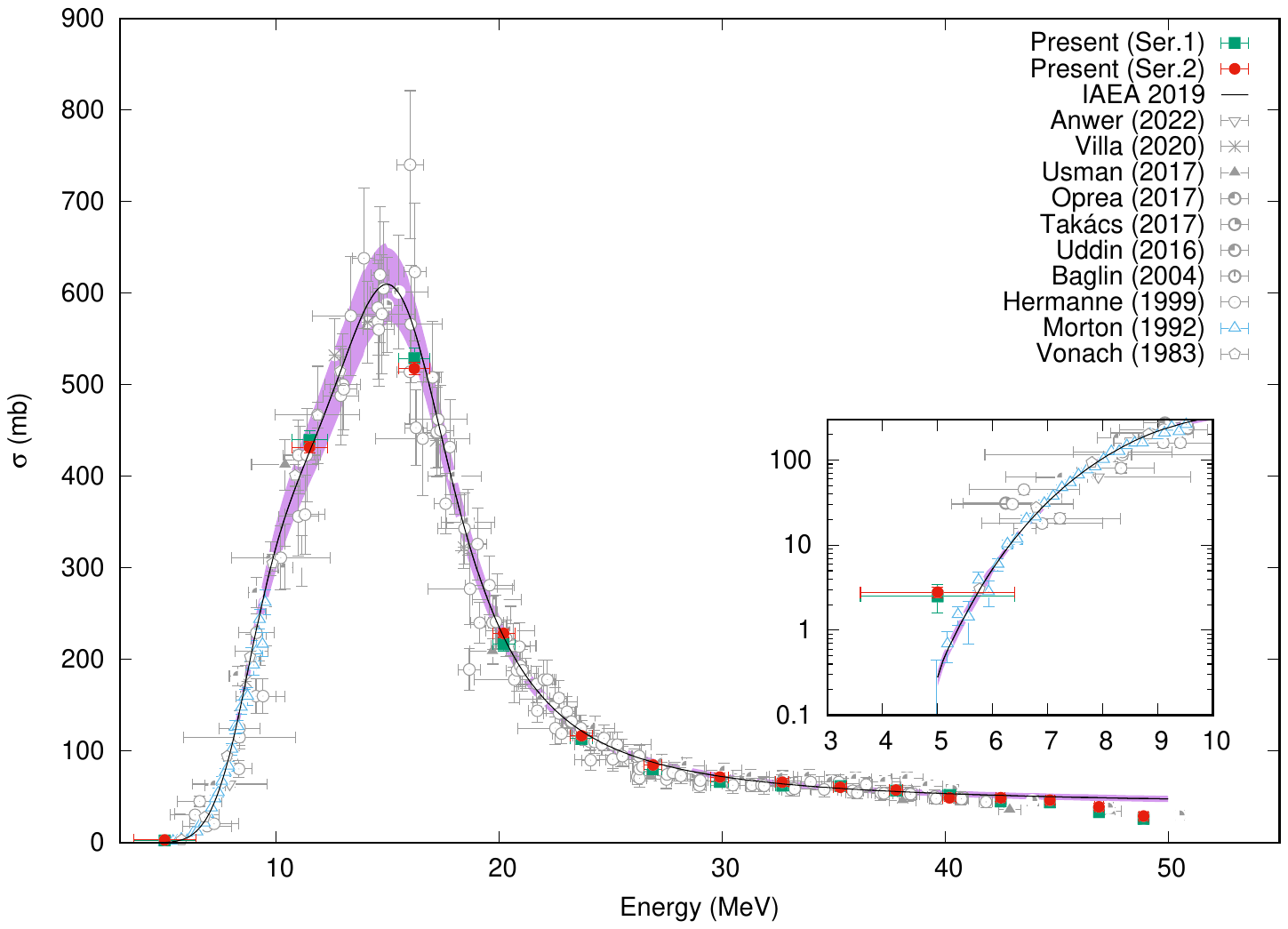}
\end{center}
\caption{
\natu Ti($\alpha$,x)$^{51}$Cr cross sections measured by us (Series 1 and 2) and recommended by the IAEA~\cite{Hermanne2018} compared with the literature data not rejected in preparation of the IAEA recommendation~\cite{Anwer2022,Villa2020,Usman2017,Oprea2017,Takacs2017,Uddin2016,Baglin2005,Hermanne1999,Morton1992,Vonach1983}.
The vertical error bars of the cross sections from the present work are for the uncertainties propagated from the uncertainties in the peak areas.
The shaded area accompanying the line indicates the uncertainty in the cross sections recommended by the IAEA.
}
\label{fig:monitor}
\end{figure}

Figure~\ref{fig:monitor} shows the comparison of the \natu Ti($\alpha$,x)$^{51}$Cr cross sections recommended by the IAEA~\cite{Hermanne2018} with those determined in the present work by the conventional activation cross section formula~\cite{Otuka2017} without decay curve analysis.
The datasets labelled by ``Ser.~1" and ``Ser.~2" were determined with the 320.1~keV peaks in the gamma spectra collected $\sim$10 to $\sim$20~h (Series 1) and $\sim$5 to $\sim$10~h” (Series 2) after irradiation, respectively.

Determination of our cross sections is based on the beam energies measured by the TOF system and beam flux estimated by the integrated charge measured by the Faraday cup.
The measured cross sections agree well with the recommended cross sections except for the lowest energy point at 5~MeV,
where the cross sections measured by us are higher than the recommended cross section following Morton \etal.'s dataset~\cite{Morton1992}
but still consistent each other within the horizontal error bars.
We accepted the measured beam flux and target thicknesses in the cross section determination without any adjustments.

\section{Simultaneous decay curve analysis}
We performed the simultaneous decay curve analysis~\cite{Otuka2024} for each Pt foil separately.
In this analysis, we model cooling time ($t$) dependence of the emission rate for the $i$th gamma peak, $A_i(t)$, by a sum of the contributions from various product nuclides and their decay paths:
\begin{equation}
A_i(t)=\left[
 \sum_j I_{ij}N_1(t,\lambda_{j1})
+\sum_k I_{ik}N_2(t,\lambda_{k1},\lambda_{k2})
+\sum_l I_{il}N_3(t,\lambda_{l1},\lambda_{l2},\lambda_{l3})
\right],
\label{eqn:emissionrate}
\end{equation}
where (1) the first term is for contribution from decay path $j$ starting from production of the nuclide $j1$,
which emits the gamma line $ij$,
(2) the second term is for contribution from a decay path $k$ starting from production of the nuclide $k1$,
which daughter $k2$ emits the gamma line $ik$,
and (3) the third term is for contribution from a decay path $l$ starting from production of the nuclide $l1$,
which granddaughter $l3$ emits the gamma line $il$.
The observed gamma peak $i$ in the spectra includes the gamma lines $ij$, $ik$ and $il$,
which may correspond to a single gamma line or several gamma lines unresolved in our measurement.
The suffixes 1, 2 and 3 of the decay constant $\lambda$ indicate the product nuclide, its daughter and granddaughter, respectively.
$I_{ij}$ is the emission probability of the gamma line $i$ for the decay path $j$.
The functions $N_1$, $N_2$ and $N_3$ are
\begin{eqnarray}
N_1(t,\lambda_{j1})&=&\phi n \sigma_{j1} g(t,\lambda_{j1})\\
N_2(t,\lambda_{k1},\lambda_{k2})&=&\phi n \sigma_{k1} p_{k1\to k2}\left[
 \frac{\lambda_{k1}}{\lambda_{k2}-\lambda_{k1}}g(t,\lambda_{k1})
+\frac{\lambda_{k1}}{\lambda_{k1}-\lambda_{k2}}g(t,\lambda_{k2})
\right]\\
N_3(t,\lambda_{l1},\lambda_{l2},\lambda_{l3})&=&\phi n \sigma_{l1} p_{l1\to l2} p_{l2\to l3}\left[
 \frac{\lambda_{l1}\lambda_{l2}}{(\lambda_{l2}-\lambda_{l1})(\lambda_{l3}-\lambda_{l1})}
+\frac{\lambda_{l1}\lambda_{l2}}{(\lambda_{l1}-\lambda_{l2})(\lambda_{l3}-\lambda_{l2})}
+\frac{\lambda_{l1}\lambda_{l2}}{(\lambda_{l1}-\lambda_{l3})(\lambda_{l2}-\lambda_{l3})}
\right],
\end{eqnarray}
where $\phi$ is the beam flux,
$n$ is the areal sample atom number density,
$\sigma_{j1}$ is the cross section for production of the nuclide $j1$,
$p_{k1\to k2}$ is the decay branching ratio of the nuclide $k1$ to $k2$,
and
\begin{equation}
g(t,\lambda)=\frac{(1-e^{-\lambda t_b})e^{-\lambda t}}{\lambda}
\end{equation}
for irradiation time $t_b$ and cooling time $t$.
In our current work, these equations covering three generations are enough.
See Appendix of our previous publication~\cite{Otuka2024} for a proof of these equations.
We assumed in our modelling that production of a nuclide with the mass number $A\le 191$ or the atomic number $Z\le 77$ is absent.~\footnote{
We searched the most intense gamma lines of the Ir and Os isotopes accessible in our experimental setup for the first Pt foil irradiated at 50.2~MeV.
We confirmed that they are absent for $^{193}$Os (30.11~d), $^{190m2}$Ir (3.087~h), $^{189}$Ir (13.2~d) and $^{188}$Ir (1.73~d).
They exist for $^{194m}$Ir (171~d), $^{194g}$Ir (19.3~h), $^{192g}$Ir (73.8~d), $^{191g}$Os (15.4~d),$^{190g}$Ir (11.8~d) and $^{185}$Os (93.6~d),
but the least-squares analysis gives unreasonable solutions for their production cross sections (\ie, negative cross sections, very large uncertainties).
Hence,
we concluded that contribution of their productions are negligible in our simultaneous decay curve analysis.
}

Table~\ref{tab:decaypath} summarizes the decay chain models of the 41 gamma lines considered in the present simultaneous decay curve analysis.
For example,
the 98~keV $\gamma$-ray emission rate modelled by 
(1) \natu Pt($\alpha$,x)$^{195}$Au and \natu Pt($\alpha$,x)$^{195m}$Pt,
(2) \natu Pt($\alpha$,x)$^{195m}$Hg$\to$$^{195}$Au and \natu Pt($\alpha$,x)$^{195g}$Hg$\to$$^{195}$Au,
and (3) \natu Pt($\alpha$,x)$^{195m}$Hg $\to$$^{195g}$Hg$\to$$^{195}$Au are expressed by the sum of the following three terms:
\begin{eqnarray*}
1&:&I_{195\mathrm{Au}}N_1(t,\lambda_{195\mathrm{Au}})+I_{195m\mathrm{Pt}}N_1(t,\lambda_{195m\mathrm{Pt}})\\
2&:&I_{195\mathrm{Au}}N_2(t,\lambda_{195m\mathrm{Hg}},\lambda_{195\mathrm{Au}})+I_{195\mathrm{Au}}N_2(t,\lambda_{195g\mathrm{Hg}},\lambda_{195\mathrm{Au}})\\
3&:&I_{195\mathrm{Au}}N_3(t,\lambda_{195m\mathrm{Hg}},\lambda_{195g\mathrm{Hg}},\lambda_{195\mathrm{Au}}),
\end{eqnarray*}
where $I_{195\mathrm{Au}}$=11.21\% and $I_{195m\mathrm{Pt}}$=11.7\% as shown in Table~\ref{tab:decaydata}.
If we have $n$ data points of gamma emission rates $\mathbf{A}=\{A_i\} (i=1,n)$ related with $m$ production cross sections $\boldsymbol{\sigma}=\{\sigma_i\} (i=1,m)$,
the production cross sections can be solved as the least-squares solution of 
\begin{equation}
\mathbf{A}=G\boldsymbol{\sigma}+\mathbf{e},
\label{eqn:lsq}
\end{equation}
where $\mathbf{A}$ and $\mathbf{e}$ (fitting residual) are $n$ dimensional vectors,
$\boldsymbol{\sigma}$ is a $m$ dimensional vector,
and $G$ is a $m\times n$ matrix collecting the coefficients relating $\mathbf{A}$ and $\boldsymbol{\sigma}$.
Note that Eq.~(\ref{eqn:lsq}) is solved for each Pt foil separately.

It is obvious from the formulation that the production cross sections determined by the simultaneous decay curve analysis are \textit{not cumulative but independent} ignoring the presence of a state with short half-life ($T_{1/2}<$ 1 min in the present work) such as the $^{196}$Au 8 sec state.

\begin{landscape}
\begin{longtable}{rccccccccccccccccccccccc}
\caption{
Contributions of reaction product nuclides to gamma lines modelled in the present work.
``1" means a gamma line emitted following $\beta$/EC/IT decay of the product nuclide (\eg, 98~keV gamma emission following \natu Pt($\alpha$,x)$^{195}$Au and its EC decay),
``2" means a gamma line emitted following $\beta$/EC/IT decay of a daughter of the product nuclide (\eg, 98~keV gamma emission following \natu Pt($\alpha$,x)$^{195g}$Hg $\to$ $^{195}$Au and its EC decay),
and ``3" means a gamma line emitted following $\beta$/EC/IT decay of a granddaughter (\eg, 98~keV gamma emission following \natu Pt($\alpha$,x)$^{195m}$Hg $\to$ $^{195g}$Hg $\to$ $^{195}$Au and its EC decay).
$E_\gamma$ is the gamma energy in keV.
The italicized gamma energy shows that the gamma line was modelled and analysed, but finally excluded from the simultaneous fitting.
Repetition of ``2" for the 186~keV gamma line due to $^{193m}$Hg production is for emission of 186.6~keV gamma-rays following $^{193m}$Hg$\to$$^{193g}$Hg and emission of 186~keV gamma-rays following $^{193m}$Hg$\to$$^{193}$Au.
}
\label{tab:decaypath}
\\
\hline
Product     &\multicolumn{2}{c}{$^{200}$Au}
            &$^{199}$Hg
            &$^{199}$Au
            &\multicolumn{2}{c}{$^{198}$Au}
            &\multicolumn{2}{c}{$^{197}$Hg}
            &\multicolumn{2}{c}{$^{197}$Pt}
            &\multicolumn{2}{c}{$^{196}$Au}
            &\multicolumn{2}{c}{$^{195}$Hg}
            &$^{195}$Au
            &$^{195}$Pt
            &$^{194}$Hg
            &$^{194}$Au
            &\multicolumn{2}{c}{$^{193}$Hg}
            &$^{192}$Hg
            &$^{192}$Au
\\
\hline
$E_\gamma$  &$m$  &$g$  &$m$  &    &$m$  &$g$  &$m$  &$g$  &$m$  &$g$  &$m$  &$g$  &$m$  &$g$  &    &$m$  &     &    &$m$   &$g$  &     &     \\
\hline                                                                                                                                        
  98        &     &     &     &    &     &     &     &     &     &     &     &     & 2,3 & 2   & 1  & 1   &     &    &      &     & 1   &     \\
\textit{133}& 1   &     &     &    &     &     & 1   &     &     &     &     &     &     &     &    &     &     &    &      &     &     &     \\
 147        &     &     &     &    &     &     &     &     &     &     & 1   &     &     &     &    &     &     &    &      &     &     &     \\
 158        &     &     & 1   & 1  &     &     &     &     &     &     &     &     &     &     &    &     &     &    &      &     & 1   &     \\
 180        &     &     &     &    & 1   &     &     &     &     &     &     &     & 2   & 1   &    &     &     &    &      &     &     &     \\
\hline                                                                                                                                        
\textit{186}&     &     &     &    &     &     &     &     &     &     &     &     &     &     &    &     &     &    & 2,2,3& 1,2 & 1   &     \\
 188        &     &     &     &    &     &     &     &     &     &     & 1   &     &     &     &    &     &     &    &      &     &     &     \\
 191        &     &     &     &    &     &     & 2   & 1   & 2   & 1   &     &     &     &     &    &     &     &    &      &     &     &     \\
 204        &     &     &     &    & 1   &     &     &     &     &     &     &     &     &     &    &     &     &    &      &     & 1   &     \\
 208        &     &     &     & 1  &     &     &     &     &     &     &     &     & 1,2 & 1   &    &     &     &    &      &     &     &     \\
\hline                                                                                                                                        
 214        &     &     &     &    & 1   &     &     &     &     &     &     &     &     &     &    &     &     &    &      &     &     &     \\
 255        & 1   &     &     &    &     &     &     &     &     &     &     &     &     &     &    &     &     &    & 2,3  & 2   &     &     \\
\textit{258}&     &     &     &    &     &     &     &     &     &     &     &     &     &     &    &     &     &    & 1,2  & 1   &     &     \\
\textit{261}&     &     &     &    &     &     &     &     &     &     &     &     & 1,2 & 1   &    &     &     &    &      &     & 1   &     \\
 274        &     &     &     &    &     &     &     &     &     &     &     &     &     &     &    &     &     &    &      &     & 1   &     \\
\hline                                                                                                                                        
 279        &     &     &     &    &     &     & 1   &     & 1   &     &     &     & 1   &     &    &     &     &    &      &     & 1   &     \\
 293        &     &     &     &    &     &     &     &     &     &     &     &     &     &     &    &     & 2   & 1  &      &     &     &     \\
 295        &     &     &     &    &     &     &     &     &     &     &     &     &     &     &    &     &     &    &      &     & 2   & 1   \\
 316        &     &     &     &    &     &     &     &     &     &     & 1   &     &     &     &    &     &     &    &      &     & 2   & 1   \\
 328        &     &     &     &    &     &     &     &     &     &     &     &     &     &     &    &     & 2   & 1  &      &     &     &     \\
\hline                                                                                                                                        
 333        & 1   &     &     &    & 1   &     &     &     &     &     & 2   & 1   &     &     &    &     &     &    &      &     &     &     \\
 346        &     &     &     &    &     &     &     &     & 1   &     &     &     &     &     &    &     &     &    &      &     &     &     \\
 355        &     &     &     &    &     &     &     &     &     &     & 2   & 1   &     &     &    &     &     &    &      &     &     &     \\
 367        & 1,2 & 1   &     &    &     &     &     &     &     &     &     &     & 1   &     &    &     &     &    &      &     &     &     \\
 374        &     &     & 1   &    &     &     &     &     &     &     &     &     &     &     &    &     &     &    &      &     &     &     \\
\hline                                                                                                                                        
\textit{381}&     &     &     &    &     &     &     &     &     &     &     &     &     &     &    &     &     &    & 1,2  & 1   &     &     \\
 387        &     &     &     &    &     &     &     &     &     &     &     &     & 1   &     &    &     &     &    &      &     &     &     \\
 411        &     &     &     &    & 2   & 1   &     &     &     &     &     &     &     &     &    &     &     &    &      &     &     &     \\
\textit{497}& 1   &     &     &    &     &     &     &     &     &     &     &     &     &     &    &     &     &    & 1    &     &     &     \\
 560        &     &     &     &    &     &     &     &     &     &     &     &     & 1   &     &    &     &     &    &      &     &     &     \\
\hline                                                                                                                                        
\textit{573}&     &     &     &    &     &     &     &     &     &     &     &     &     &     &    &     &     &    & 1    &     &     &     \\
\textit{579}& 1   &     &     &    &     &     &     &     &     &     &     &     &     &     &    &     &     &    & 2    & 1   &     &     \\
 585        &     &     &     &    &     &     &     &     &     &     &     &     & 2   & 1   &    &     &     &    &      &     &     &     \\
 599        &     &     &     &    &     &     &     &     &     &     &     &     & 2   & 1   &    &     &     &    &      &     &     &     \\
 759        & 1   &     &     &    &     &     &     &     &     &     &     &     &     &     &    &     &     &    &      &     &     & 1   \\
\hline                                                                                                                                        
 779        &     &     &     &    &     &     &     &     &     &     &     &     & 2   & 1   &    &     &     &    &      &     &     &     \\
 861        &     &     &     &    &     &     &     &     &     &     &     &     &     &     &    &     &     &    & 2    & 1   &     &     \\
\textit{932}&     &     &     &    &     &     &     &     &     &     &     &     & 2   & 1   &    &     & 2   & 1  & 1    &     &     &     \\
1111        &     &     &     &    &     &     &     &     &     &     &     &     & 2   & 1   &    &     &     &    &      &     &     &     \\
1118        &     &     &     &    &     &     &     &     &     &     &     &     &     &     &    &     &     &    & 2    & 1   &     &     \\
\hline                                                                                                                                        
1468        &     &     &     &    &     &     &     &     &     &     &     &     &     &     &    &     & 2   & 1  &      &     &     &     \\
\hline
\end{longtable}
\end{landscape}

Among the product nuclides considered,
we could not include the gamma-rays emitted directly from $^{194}$Hg (no gamma-ray emission) and $^{193m}$Hg (incomplete knowledge of intensities for gamma-rays emitted after its EC/$\beta^+$ decay).
The $^{194}$Hg yield was quantified by analysis of the gamma-rays emitted from the daughter nuclide $^{194}$Au,
and $^{193m}$Hg production was quantified by analysis of the gamma-rays emitted from the daughter nuclide $^{193g}$Hg and granddaughter nuclide $^{193}$Au.

For $^{193m}$Hg,
the absolute probabilities of gamma-ray emission following its EC/$\beta^+$ decay are not determined in the latest ENSDF evaluation~\cite{A193} due to difficulty in separation of $^{193m}$Hg and $^{193g}$Hg decays for the gamma lines shared by both decays.
Due to this situation,
we excluded all gamma lines associated with $^{193m}$Hg EC/$\beta^+$ decay and other gamma lines overlapping with them\footnote{
They are $^{200m}$Au 497.8~keV, $^{195g}$Hg 930.9~keV, $^{193g}$Hg 258.0~keV and 381.6~keV gamma lines.
}
in our analysis.
Such gamma lines are italicized in Tables~\ref{tab:decaydata} and \ref{tab:decaypath}.
It allowed us to use only decay gamma lines of $^{193g}$Hg free from interference by $^{193m}$Hg EC/$\beta^+$ decay gammas.
%
%

We observed unexpected finite count rates ($\sim$1 decay/sec) for the 133~keV gamma line in the last measurement ($\sim$60 days after irradiation) of several Pt foils.
The simultaneous fitting result is not influenced so much by inclusion or exclusion of the 133~keV gamma line,
and we excluded this gamma line ($^{200m}$Au 133.2~keV, $^{197m}$Hg 134.0~keV) from the data reduction process to be on the safe side.
The origin of the 133~keV gamma line peak area after long cooling could be attributed to coincidence summing effect caused by emission of two X-rays.
We also excluded (1) the 261~keV gamma line ($^{195m,g}$Hg 261.8~keV, $^{192}$Hg 262.6~keV) because it often introduced large chi-square, and (2) the 579~keV gamma line ($^{200m}$Au 579.3~keV, $^{193g}$Hg 581.0~keV) since the decay curve of this gamma line from a thick Pt target irradiated by 50~MeV alpha-particle by us (unpublished) shows unknown long half-life component.

We fixed the production cross section to zero when
(1) the incident energy is lower than the threshold energy,
(2) the strongest gamma line of the product is not observed,
(3) the cross section becomes negative if included in fitting, or
(4) the cross section uncertainty becomes larger than 100\% if included in fitting.
We also excluded from fitting four measured emission rates which obviously do not follow the trend.

\section{Results}
\subsection{Decay curves}
\begin{figure}[hbtp]
\begin{center}
\includegraphics[width=0.8\textwidth]{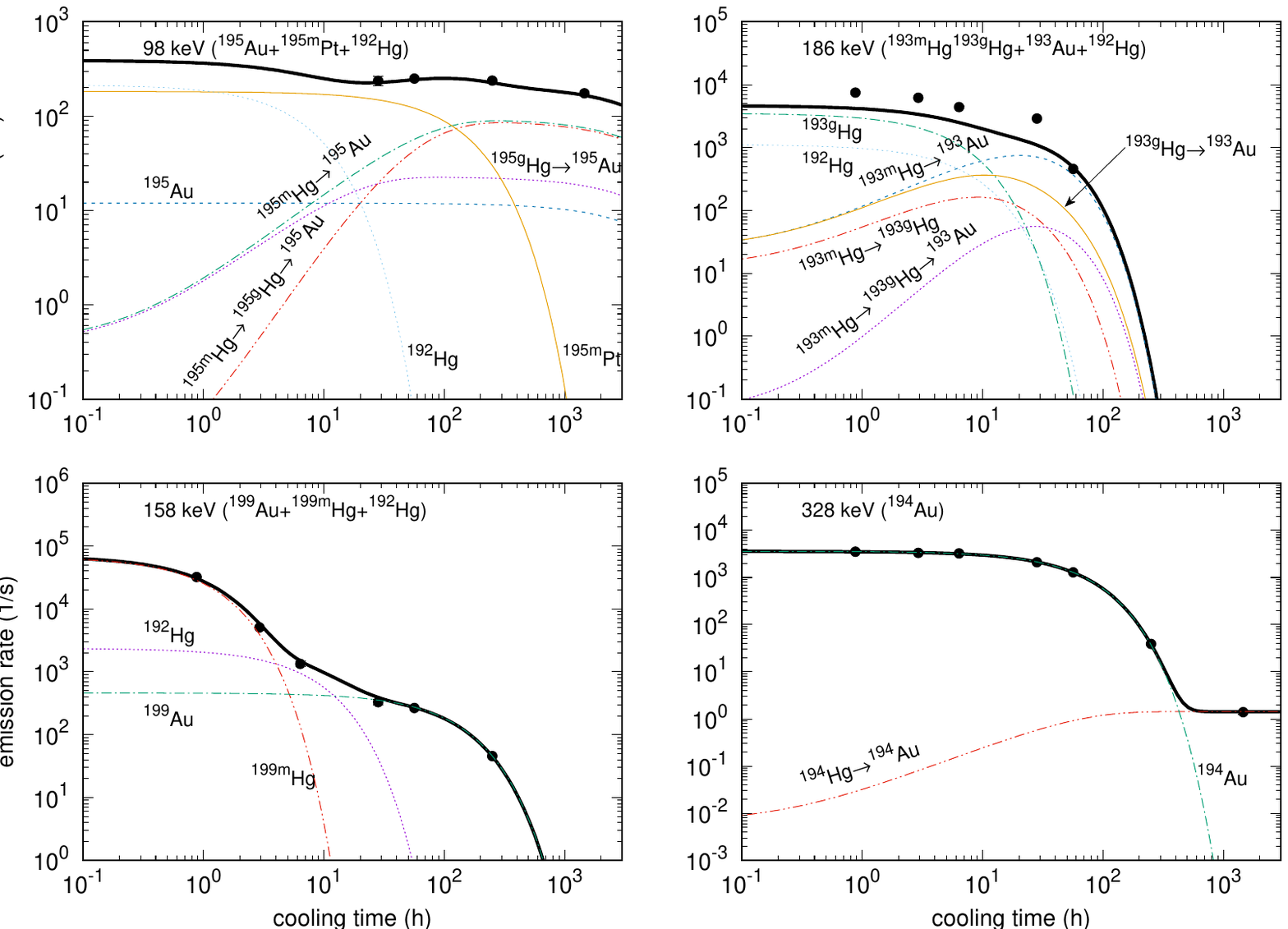}
\end{center}
\caption{
Emission rates of four gamma lines and their decompositions to the contributions from various reaction products and decay paths for the Pt foil irradiated at 50.2~MeV.
The parenthesized nuclide symbols indicate the nuclides emitting the gamma line. The error bars of the measured rates originate from the uncertainties in the peak areas though they are almost invisible.
}
\label{fig:decaycurve}
\end{figure}

Figure~\ref{fig:decaycurve} shows examples of the gamma emission rates from measurements of the Pt foil placed at the most upstream side of the stack.
In each panel of this figure,
the total decay curve (solid line) and partial decay curves (dashed lines) correspond to the left- and right-hand sides of Eq.~(\ref{eqn:emissionrate}), respectively,
and they are adjusted to reproduce the measured emission rates (solid symbols) by solving Eq.~(\ref{eqn:lsq}).
The measured emission rates are corrected for the detection efficiency and gamma self-absorption.
The parenthesized nuclide symbols show the nuclides emitting the gamma-rays.
They may be one directly produced by a reaction (``1" in Table~\ref{tab:decaypath}), or a decay product of another reaction product (``2" or ``3" in Table~\ref{tab:decaypath}).
\begin{itemize}
\item
98 keV: All partial emission rates other than the $^{192}$Hg one originate from the same transition in $^{195}$Pt from its first excitation level to the ground state. 
This plot shows that three curves originating from $^{195m}$Hg and $^{195g}$Hg productions follow the half-life of $^{195}$Au after long cooling ($\sim$100~h) as seen in our previous work~\cite{Otuka2024}.
\item
158 keV: The emission rates from the first measurement and last three measurements represent $^{199m}$Hg and $^{199}$Au production, respectively,
and they allow use of the conventional activation cross section formula without decay curve analysis for determination of their production cross sections.
\item
186 keV: We considered six decay paths contributing to the measured emission rates.
Unfortunately, their sum still underestimates the measured emission rates except for the one from the last measurement,
and we excluded this gamma line ($^{193g}$Hg 186.6~keV, $^{193}$Au 186.2+187.8~keV, $^{192}$Hg 186.4~keV) in the cross section determination.
\item
328 keV: The emission rate measured after the longest cooling time is explained by decay of $^{194}$Au following decay of the long lived $^{194}$Hg (447~y).
We will show that the \natu Pt($\alpha$,x)$^{194}$Hg cross section (not seen in the literature) has a large value ($\sim$600 mb at 50~MeV).
\end{itemize}

Table~\ref{tab:chi2} summarizes the degree of freedom (number of the measured emission rate data points used in fitting minus the number of the nuclides which production cross sections were determined by fitting\footnote{
For example,
the degree of freedom is 4=6-2 for the last (16th) foil for which \underline{6} measured emission rates (5 for the 98~keV gamma line and 1 for the 158~keV gamma line) were usable to determine \underline{2} cross sections for production of $^{199m}$Hg and $^{195m}$Pt.
The cross sections for production of other nuclides ($^{199}$Au, $^{195m,g}$Hg, $^{195}$Au, $^{192}Hg$) were set to zero for this Pt foil because the alpha-particle energy is below the threshold energies for their productions.
}
) and the reduced chi-square (chi-square divided by the degree of freedom) for each Pt foil.
The reduced chi-square becomes 1 if the best consistency within error bars is achieved between the measured emission rates and the decay curves obtained by fitting,
and it becomes higher if there is inconsistency beyond the error bars.
This table shows the reduced chi-square is higher than 1 for all Pt foils.
There are various possible reasons of this inconsistency (\eg, statistical fluctuation, overlooked interference gammas, error in decay data adopted).
This implies that the cross section determination based on a single measurement without following the decay profile could suffer from these defects.
The uncertainties of the excitation functions shown in the figures and table of this article are external uncertainties,
namely the uncertainties obtained as the least-squares solution multiplied by the square root of the reduced chi-square to achieve better consistency between the measured emission rates and fitting result (\ie, $\chi^2_\textrm{red}$=1).
It means that the uncertainties in the figures and tables are from the uncertainties in the peak areas and from the fitting procedure. 

\begin{table}
\caption{
Degree of freedom, reduced chi-square ($\chi^2_\textrm{red}$) and its square root at each Pt foil.
}
\label{tab:chi2}
\begin{tabular}{lrrrrrrrrrrrrrrrr}
\hline
Pt foil                             &  1   &    2  &  3   &  4    &  5   &  6   &   7   &   8   &   9   &  10  &  11  &  12  &  13   &  14  &  15  &  16  \\
$E$ (MeV)                           & 50.2 &  48.1 & 46.0 & 43.9  & 41.6 & 39.3 & 36.8  & 34.3  & 31.6  & 28.8 & 25.7 & 22.4 & 18.7  & 14.5 &  9.3 &  2.2 \\
\hline                               
Degree of freedom                   &  134 &   134 &  126 &  119  &  111 &  104 &  102  &   92  &   94  &   78 &   31 &   44 &   24  &   15 &   17 &    4 \\
$\chi^2_\textrm{red}$               & 22.7 &  23.6 &  5.8 & 10.8  &  6.9 & 12.5 &  8.1  &  5.3  &  8.4  &  5.9 &  4.5 &  5.6 &  2.4  &  1.2 & 12.2 & 40.9 \\
Square root of $\chi^2_\textrm{red}$&  4.8 &   4.9 &  2.4 &  3.3  &  2.6 &  3.5 &  2.8  &  2.3  &  2.9  &  2.4 &  2.1 &  2.4 &  1.6  &  1.1 &  3.5 &  6.4 \\

\hline
\end{tabular}
\end{table}

\subsection{Excitation functions for production of $^{198}$Au, $^{197}$Hg and $^{195}$Hg}
In our previous work~\cite{Otuka2024},
we found that the excitation functions of $^{198g+m}$Au, $^{197g+m}$Hg and $^{195g+m}$Hg productions up to 29~MeV measured by us are consistent with those from simulations by TALYS-2.0 in general, and the agreement for the isomeric ratio is improved if we reduce the spin cut-off parameter $\eta$ from 1.
In the present work,
we extended the energy range of our analysis to 50~MeV as shown in Figs.~\ref{fig:198Au} to \ref{fig:195Hg}.
Hereafter,
the sum of the ground state and metastable state production cross sections $\sigma_{g+m}=\sigma_g+\sigma_m$ is quoted as the total production cross section.
The cross sections published by Capurro \etal.~\cite{Capurro1991} are the elemental cross sections (cross section for a natural sample) divided by the atomic mass of natural platinum (see~\cite{Bonesso2017}),
and their cross sections converted back to the elemental cross sections by us are shown in these figures.
It has been known that the production cross sections from the measurement by Sagaidak \etal.~\cite{Sagaidak1979} are systematically higher.
Following Hermanne \etal.'s comment on Sagaidak \etal.'s $^{193}$Hg production cross sections~\cite{Hermanne2006},
we interpret the cross sections published by Sagaidak \etal. as the elemental cross sections divided by the sum of the natural isotopic abundances of the Pt target isotopes contributing to the production listed in Table 1 of~\cite{Sagaidak1979},
and we present their cross sections converted to the elemental cross sections in the figures.
The numerical data from our new measurement are tabulated in Appendix.

\subsubsection{$^{198}$Au}
Figure~\ref{fig:198Au} compares the $^{198m,g,g+m}$Au production cross sections and the isomeric ratio $\sigma_m/\sigma_{g+m}$ with those from the previous measurements~\cite{Otuka2024,Hermanne2006,Capurro1991,Sagaidak1979} and from the TALYS-2.0 simulation.
The newly measured excitation functions show excellent agreement with those measured by us previously and Hermanne \etal. The total production cross section reported by Sagaidak \etal. is multiplied by $\sim$0.59 (sum of the natural isotopic abundances of $^{195}$Pt and $^{196}$Pt) following Hermanne \etal.'s interpretation but still higher than those from other measurements.
The half-lives of $^{198m}$Au and $^{198g}$Au are close to each other,
and one can expect the cumulative production cross section expressed by
\begin{equation}
\sigma_g^\mathrm{cum}=\sigma_g + \frac{\lambda_m}{\lambda_m - \lambda_g}\sigma_m
\label{eqn:cumxs}
\end{equation}
\cite{Otuka2024} is much higher than the total production cross section because $\lambda_m / (\lambda_m-\lambda_g)\sim 6.4$ for $^{198m,g}$Au.
We constructed the cumulative production cross section from $\sigma_g$ and $\sigma_m$ measured by us.
Figure~\ref{fig:198Au} shows that our cumulative production cross section agrees with the cross section reported by Sagaidak \etal. multiplied by 0.59,
which indicates that the cross section reported by Sagaidak \etal. is the one defined by Eq.~(\ref{eqn:cumxs}).
This is a good example demonstrating that the cumulative cross section determined by the ground state activity after long cooling may overestimate the total production cross section.
The TALYS-2.0 simulation estimates the energy dependence of the excitation function for the total production cross section very well.
We also see that the simulation result for isomer productions is very sensitive to the spin cutoff parameter.
The isomeric ratio measured by us prefers reduction of the spin cutoff parameter from the rigid body model value ($\eta$=1).
Above 45~MeV,
our total production cross section is consistent with the one measured by Capurro \etal.,
but only our measurement shows increase of $\sigma_m$ and $\sigma_m/\sigma_{g+m}$,
and the TALYS simulation does not support this trend.

\begin{figure}[hbtp]
\begin{center}
\includegraphics[width=0.8\textwidth]{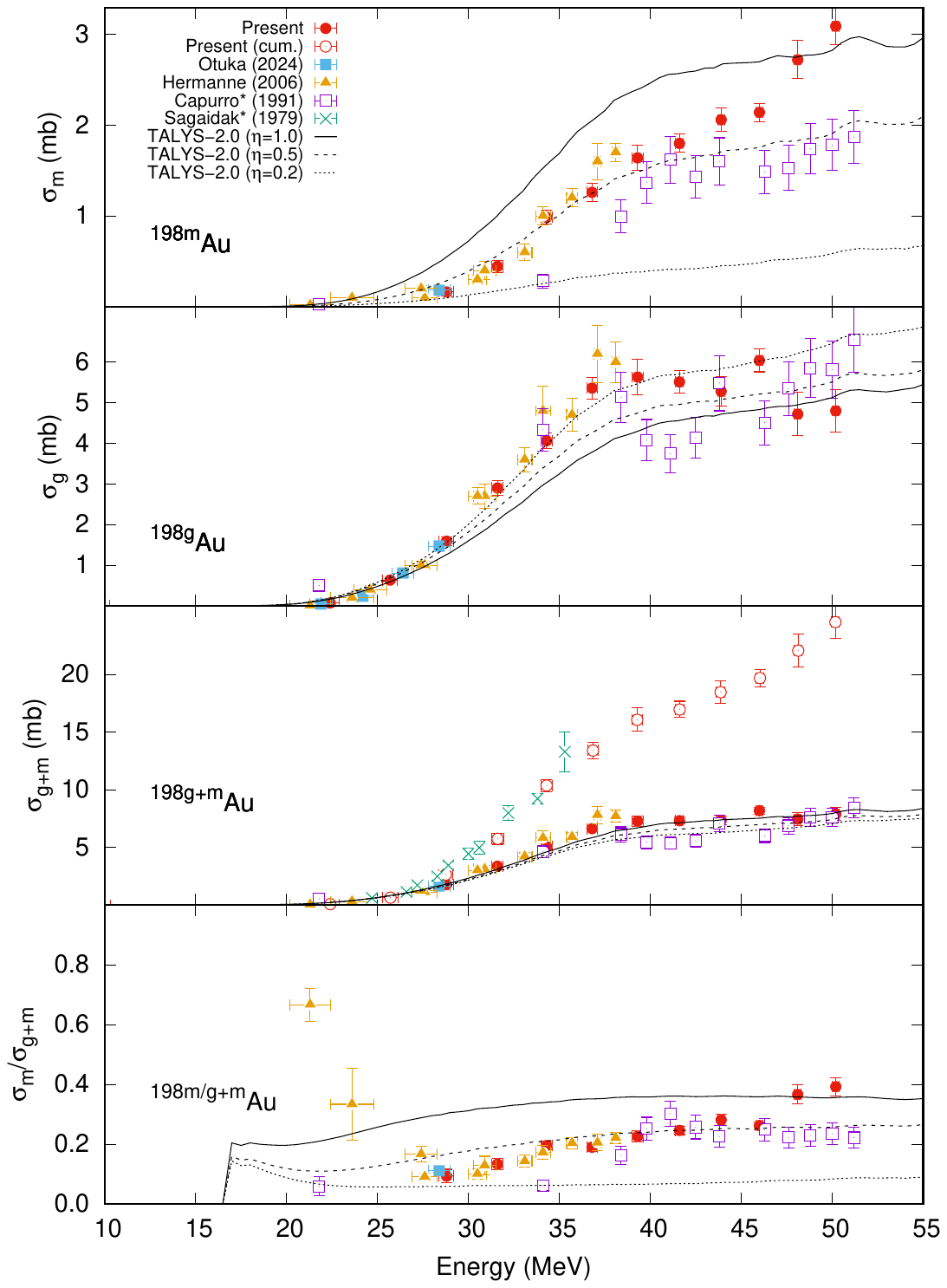}
\end{center}
\caption{
\natu Pt($\alpha$,x)$^{198m,g}$Au cross sections along with their sum $\sigma_{g+m}$ and ratio $\sigma_m/\sigma_{g+m}$.
Present (cum.) is $\sigma_g + [\lambda_m/(\lambda_m-\lambda_g)] \sigma_m$ derived from $\sigma_g$ and $\sigma_m$ determined in the present work.
The cross sections reported by Capurro \etal.~\cite{Capurro1991} are multiplied by the atomic mass of natural platinum.
The cross section reported by Sagaidak \etal.~\cite{Sagaidak1979} is multiplied by the sum of the natural isotopic abundances of $^{195}$Pt and $^{196}$Pt.
The $^{198g+m}$Au production cross sections and isomeric ratios of Hermanne \etal.~\cite{Hermanne2006} and Capurro \etal. were derived by us from their isomer production cross sections.
}
\label{fig:198Au}
\end{figure}

\subsubsection{$^{197}$Hg}
\begin{figure}[hbtp]
\begin{center}
\includegraphics[width=0.8\textwidth]{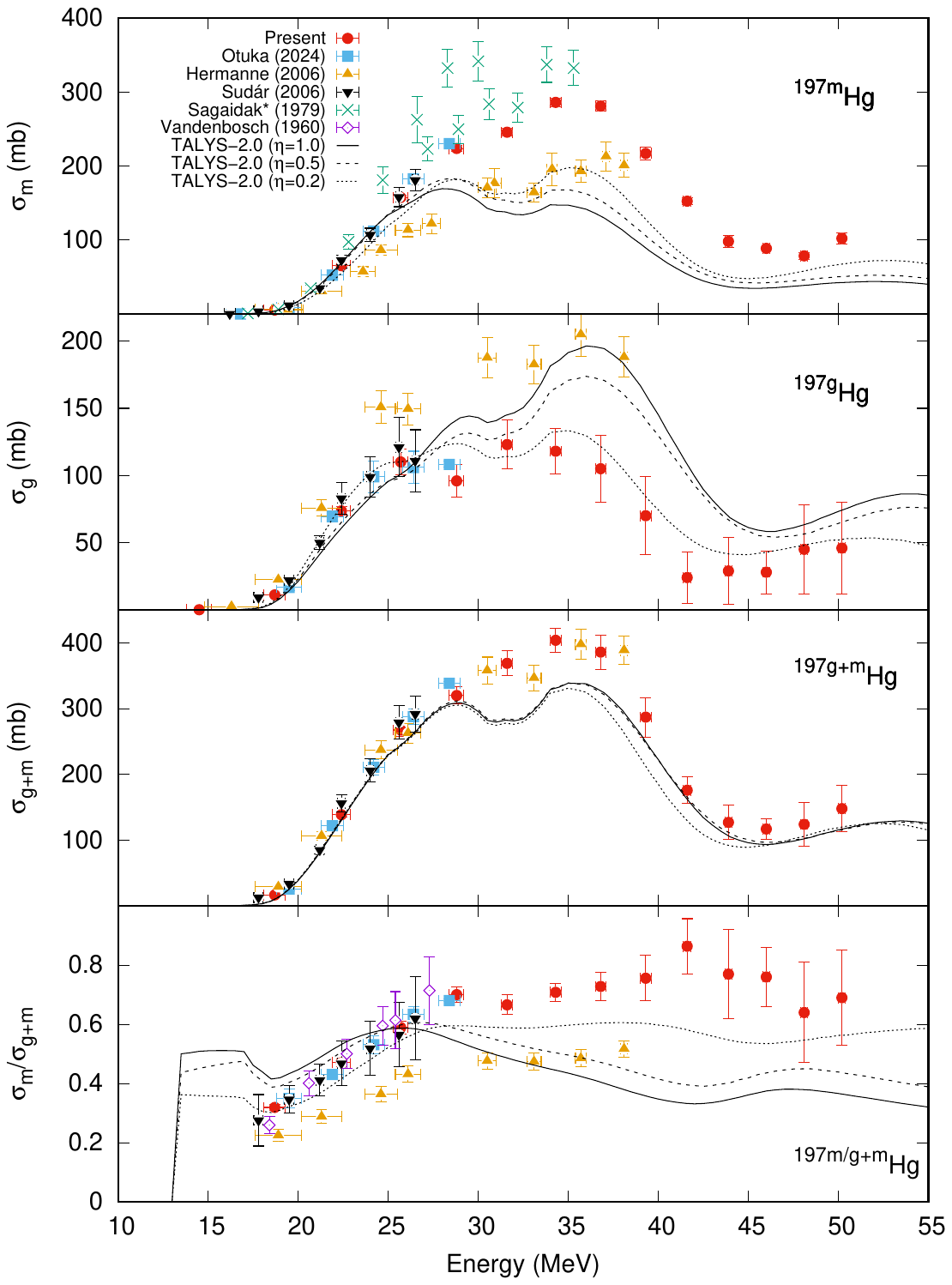}
\end{center}
\caption{
\natu Pt($\alpha$,x)$^{197m,g}$Hg cross sections along with their sum $\sigma_{g+m}$ and ratio $\sigma_m/\sigma_{g+m}$.
The cross sections of Sagaidak \etal.~\cite{Sagaidak1979} are multiplied by the sum of the natural isotopic abundances of $^{194}$Pt, $^{195}$Pt and $^{196}$Pt.
The $^{197g+m}$Hg production cross sections of Hermanne \etal.~\cite{Hermanne2006} and Sud{\'a}r \etal.~\cite{Sudar2006} as well as the isomeric ratio of Hermanne \etal. were derived by us from their isomer production cross sections.
}
\label{fig:197Hg}
\end{figure}

Figure~\ref{fig:197Hg} compares the $^{197m,g,g+m}$Hg production cross sections and the isomeric ratio $\sigma_m/\sigma_{g+m}$ from the new measurement with those in the literature~\cite{Otuka2024,Hermanne2006,Sudar2006,Sagaidak1979,Vandenbosch1960} and from the TALYS-2.0 simulation.
The newly measured excitation functions show excellent agreement with those measured by us previously as well as by Sud{\'a}r \etal. and Vandenbosch \etal.
The metastable state production cross sections measured by us and those reported by Sagaidak \etal. become close to each other if we multiply Sagaidak \etal.'s cross section by the sum of the natural isotopic abundances of $^{194}$Pt, $^{195}$Pt and $^{196}$Pt ($\sim$0.92).
The total production cross section published by Hermanne \etal. agrees with the one newly determined by us,
but the isomeric ratio determined by Hermanne \etal. is systematically lower than our result.
%
Sud{\'a}r \etal. and Vandenbosch \etal. used only the 134~keV gamma line for quantification of the $^{197m}$Hg production,
while Hermanne \etal. adopt both 134~keV and 279~keV gamma lines.
As previously discussed by us~\cite{Otuka2024} and Lebeda \etal.~\cite{Lebeda2020},
the gamma emission probability of the 279~keV gamma line seen in the latest ENSDF evaluation~\cite{A197} and adopted by Hermanne \etal. (6.1\%) should be around 3.9\%.
This explains why the $^{197m}$Hg production cross section and isomeric ratio obtained by Hermanne \etal. are lower than the cross section determined by us.

The TALYS-2.0 simulation describes the total production cross section in the measured energy region very well except for the peak region around 32~MeV,
where TALYS-2.0 shows two peaks due to $^{195}$Pt($\alpha$,2n)$^{197g+m}$Hg and $^{196}$Pt($\alpha$,3n)$^{197g+m}$Hg reactions separately.
The figure shows adjustment of the spin cutoff parameter does not improve description of the isomeric ratios measured by neither us nor Hermanne \etal. above 30~MeV.
Sud{\'a}r \etal. demonstrates that the STAPRE~\cite{Uhl1976} code describes the isomeric ratio measured by them successfully with $\eta$=0.20,
but their discussion covers the energy range up to 28~MeV only.

\subsubsection{$^{195}$Hg}
\begin{figure}[hbtp]
\begin{center}
\includegraphics[width=0.8\textwidth]{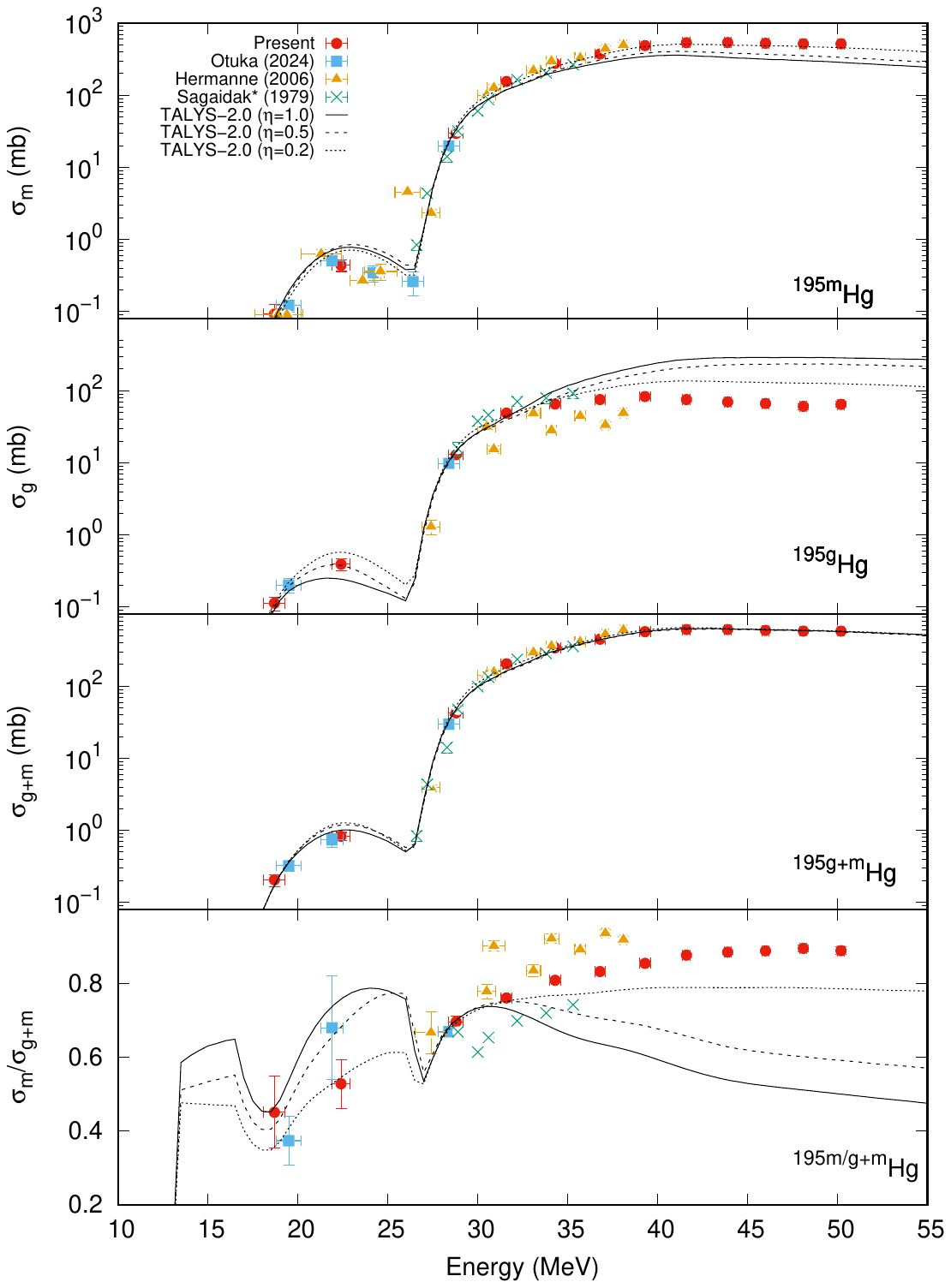}
\end{center}
\caption{
\natu Pt($\alpha$,x)$^{195m,g}$Hg cross sections along with their sum $\sigma_{g+m}$ and ratio $\sigma_m/\sigma_{g+m}$.
The cross sections of Sagaidak \etal.~\cite{Sagaidak1979} are multiplied by the sum of the natural isotopic abundances of $^{192}$Pt and $^{194}$Pt.
The $^{195g+m}$Hg production cross sections and isomeric ratios of Hermanne \etal.~\cite{Hermanne2006} and Sagaidak \etal. were derived by us from their isomer production cross sections.
}
\label{fig:195Hg}
\end{figure}

Figure~\ref{fig:195Hg} compares the $^{195m,g,g+m}$Hg production cross sections and the isomeric ratio $\sigma_m/\sigma_{g+m}$ with the previous measurements~\cite{Otuka2024,Hermanne2006,Sagaidak1979} and the TALYS-2.0 simulation results.
The newly measured excitation functions agree very well with those measured by us previously.
The gamma emission probabilities of the metastable state adopted by us are about 10\% lower than those in the latest ENSDF evaluation~\cite{A195} as discussed in our previous publication~\cite{Otuka2024}.
The ground state production cross section measured by Hermanne \etal. is slightly lower than ours,
and it makes their isomeric ratio higher than ours.
As we have already seen for $^{198}$Au and $^{197}$Hg,
the cross sections measured by us and reported by Sagaidak \etal. become close to each other if we multiply Sagaidak \etal.'s cross section by the sum of the natural isotopic abundances of $^{192}$Pt and $^{194}$Pt ($\sim$0.34).

The TALYS-2.0 simulation describes the total production cross section satisfactory in the energy region covered in the present work.
From our previous measurement up to 29~MeV,
we got only three data points of the isomeric ratio without smooth energy dependence,
and it was not clear whether reduction of the spin cutoff parameter helps improve the simulation performance.
After addition of the new data points extended to 50~MeV,
it is clear that our new measurement supports the result with $\eta$=0.2 among the three curves from the TALYS-2.0 simulation.
Note that the TALYS-2.0 curves for $\sigma_g$ are wrongly repeated as $\sigma_m$ in the figure of our previous publication (Figure 7 of~\cite{Otuka2024}).

\subsection{Excitation functions of other products}
\begin{figure}[hbtp]
\begin{center}
\includegraphics[width=0.8\textwidth]{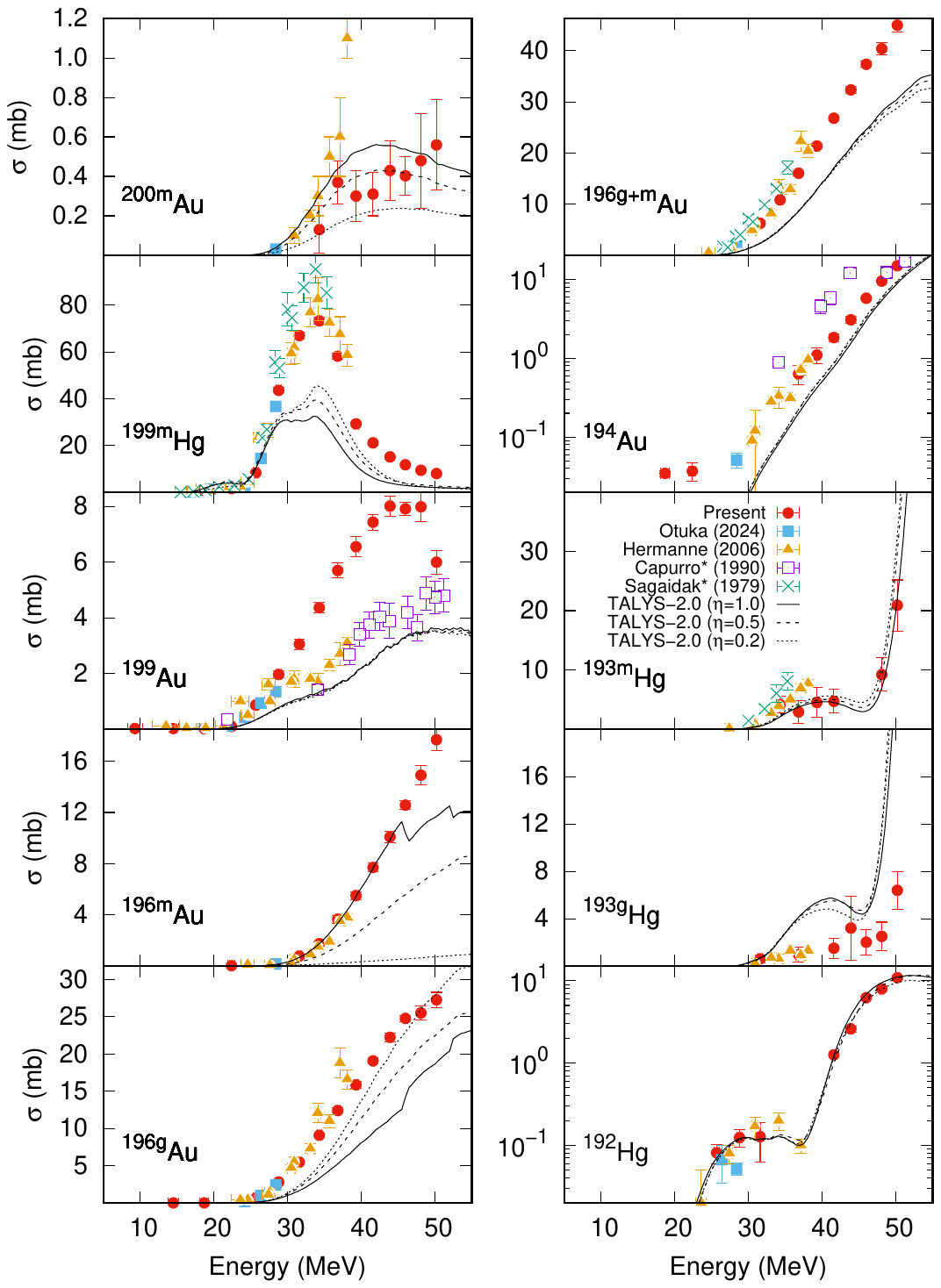}
\end{center}
\caption{
\natu Pt($\alpha$,x)$^{200m}$Au, $^{199m}$Hg, $^{199}$Au, $^{196m,g,g+m}$Au, $^{194}$Au, $^{193m,g}$Hg and $^{192}$Hg cross sections.
The cross sections reported by Capurro \etal.~\cite{Capurro1991} are multiplied by the atomic mass of natural platinum.
The cross sections of Sagaidak \etal.~\cite{Sagaidak1979} are multiplied by the sum of the natural isotopic abundances of the Pt target isotopes contributing to the production (see the main text for the Pt isotopes considered).
The $^{196g+m}$Au production cross sections of Hermanne \etal.~\cite{Hermanne2006} were derived by us from their isomer production cross sections.
}
\label{fig:others}
\end{figure}

Figure~\ref{fig:others} shows the excitation functions of some other products which we can compare with the measurements by other groups.
The numerical data from our new measurement are tabulated in Appendix.

\subsubsection{$^{200m}$Au}
The newly measured cross section shows a smooth connection with the cross section determined by us previously~\cite{Otuka2024}.
The excitation function measured by Hermanne \etal.~\cite{Hermanne2006} shows different energy dependence.
The gamma emission probabilities adopted by them are about 10\% lower than those adopted by us,
but it is not enough to explain the large difference in the cross sections from the two measurements.

\subsubsection{$^{199m}$Hg}
The newly measured cross section is consistent with the cross section determined by us previously~\cite{Otuka2024} and also the one published by Hermanne \etal.~\cite{Hermanne2006}.
Furthermore,
the cross section reported by Sagaidak \etal.~\cite{Sagaidak1979} multiplied by the sum of the natural isotopic abundances of $^{196}$Pt and $^{198}$Pt ($\sim$0.33) also become consistent with our cross sections.
The ground state is stable, and activation measurements cannot determine the isomeric ratio.

\subsubsection{$^{199}$Au}
The newly measured cross section is consistent with the one determined by us previously~\cite{Otuka2024}.
Above 30~MeV, our cross section is much higher than the cross sections published by Hermanne \etal.~\cite{Hermanne2006} and Capurro \etal.~\cite{Capurro1991}.
Our measurement and Hermanne \etal.'s measurement adopt the 158~keV gamma line ($I$=40\%) and 208~keV gamma line ($I$=8.72\%) while Capurro \etal. adopt only the 158~keV gamma line ($I$=36.9\%).
The 158~keV gamma line associated with the transition from the first excitation level of $^{199}$Hg (5/2$^-$) to the ground state (1/2$^-$) is also emitted in decay of $^{199m}$Hg.
Therefore,
one should utilize the peak area of this gamma line measured after complete decay of $^{199m}$Hg (42.67~min) for quantification of $^{199}$Au (3.139~d) production as demonstrated in Fig.~\ref{fig:decaycurve} if it is done without decay curve analysis.
But this does not explain the discrepancy seen in the measurements.

\subsubsection{$^{196m,g}$Au}
Our total production cross section is consistent with the one reported by Sagaidak \etal.~\cite{Sagaidak1979} multiplied by the sum of the natural isotopic abundances of $^{194}$Pt and $^{195}$Pt ($\sim$0.66).
The TALYS-2.0 simulation with the default spin cutoff parameter ($\eta$=1.0) seems very successful in estimation of the $^{196m}$Au production cross section.
However, it underestimates the total production cross section.
Namely, adjustment of the spin cutoff parameter does not improve the overall description.

\subsubsection{$^{194}$Au}
The newly measured production cross section has a smooth connection with the one measured by us previously~\cite{Otuka2024},
and they also agree very well with the one measured by Hermanne \etal.~\cite{Hermanne2006}.
On the other hand, the cross section reported by Capurro \etal.~\cite{Capurro1991} is systematically high.
Figure~\ref{fig:decaycurve} shows quantification of the $^{194}$Au by peak areas of the 328~keV gamma line after cooling longer than $\sim$600~h is influenced by decay of the long-lived $^{194}$Hg ($T_{1/2}$=447~y).
However, Capurro \etal. mention their measurement was done after short ($\sim$220~h) cooling.
Therefore, we consider the higher cross section reported by Capurro \etal. is not due to extra contribution of the $^{194}$Hg production.

\subsubsection{$^{193m,g}$Hg}
The threshold energies for $^{193m,g}$Hg production reactions are around 12~MeV, but we could not determine their cross sections in our previous measurement up to 29~MeV.
The newly measured isomer production cross sections are consistent with those measured by Hermanne \etal.~\cite{Hermanne2006}.
The metastable state production cross section reported by Sagaidak \etal.~\cite{Sagaidak1979} becomes closer to ours after multiplication of their cross section by the sum of the natural isotopic abundances of $^{190}$Pt and $^{192}$Pt ($\sim$0.0079) though their cross section is still higher.

As discussed in the relation with Table~\ref{tab:decaydata},
an overall normalisation factor of the relative gamma emission probabilities for the EC/$\beta^+$ branch of $^{193m}$Hg is not given in the latest ENSDF library due to difficulty in separation of the decays from $^{193m}$Hg and $^{193g}$Hg~\cite{A193}.
The normalisation factor was determined in the previous ENSDF evaluation~\cite{Achterberg2006}, but an irradiation of a thick Pt foil performed by us in 2024 (unpublished) shows inconsistency in the thick target yields determined with this normalisation factor for major gamma lines associated with the EC/$\beta^+$ branch of the $^{193m}$Hg decay.
The gamma-rays emitted in the isomeric transition are free from presence of $^{193g}$Hg, but there are only two gamma lines associated with the isomeric transition at 39.5~keV ($I$=0.3\%) and 101.3~keV ($I$=0.001\%), which are not suitable for analysis within our experimental capability.
Due to this situation, our quantification of $^{193m}$Hg production had to rely on the gamma-ray emission following (1) decay of $^{193g}$Hg free from interference with decay of $^{193m}$Hg and (2) decay of the daughter product $^{193}$Au.
Table~\ref{tab:decaydata} shows one of the most intense gamma lines of $^{193g}$Hg and $^{193}$Au are very close (186.6~keV and 186.2+187.8~keV) and inseparable, and use of the decay curve analysis seems very useful.
However, our model for 186~keV gamma emission does not fit to the measured emission rates well as shown in Fig.~\ref{fig:decaycurve}, and we excluded this gamma line from the present analysis.
As shown in Table~\ref{tab:decaydata},
we finally determined the $^{193m,g}$Hg production cross sections by 861.1 and 1118.8~keV gamma lines of $^{193g}$Hg and 255.6~keV gamma lines of $^{193}$Au.

To check possible contribution of \natu Pt($\alpha$,x)$^{193}$Au to the measured 255.6~keV peak areas,
we attempted fitting including direct production of $^{193}$Au for the first and second Pt foils with the $^{193m,g}$Hg production cross sections determined by us as their initial values,
and found that it does not show any sensitivity to presence of the direct $^{193}$Au production,
namely no change in the initial values of the $^{193m,g}$Hg and $^{193}$Au production cross sections but with large uncertainty in the $^{193}$Au production cross section after fitting.
We also tried fitting with a same initial value to the $^{193m,g}$Hg and $^{193}$Au production cross sections,
but it showed either a negative value or large uncertainty for $^{193m}$Hg and $^{193}$Au production cross sections with large correlation (more than 90\%) between them.
Based on these results,
we concluded the \natu Pt($\alpha$,x)$^{193}$Au cross section is negligible.
We observed finite \natu Pt($\alpha$,x)$^{192}$Au cross section of about 1~mb above 40~MeV,
and we cannot exclude the possibility to have small fraction of \natu Pt($\alpha$,x)$^{193}$Au cross section in the \natu Pt($\alpha$,x)$^{193m,g}$Hg cross sections determined by us.
It would be possible to determine the production cross sections of these three nuclides directly
if the overall normalisation factor of the gamma lines associated with $^{193m}$Hg EC/$\beta^+$ decay becomes known.

The TALYS-2.0 simulation reproduces our excitation function of $^{193m}$Hg production very nicely with the default spin cutoff parameter ($\eta$=1.0).
However, it overestimates the excitation function of $^{193g}$Hg production even if we reduce its production by changing the spin cutoff parameter to $\eta$=0.2.

\subsubsection{$^{192}$Hg}
The newly measured cross section is consistent with one of the two data points determined by us previously~\cite{Otuka2024} and also the one published by Hermanne \etal.~\cite{Hermanne2006}.
The TALYS-2.0 simulation also successfully reproduces the measured excitation function over the energy range investigated in the present work.

\subsection{$^{195m}$Pt}
\label{sec:195mPt}
\begin{figure}[hbtp]
\begin{center}
\includegraphics[width=0.8\textwidth]{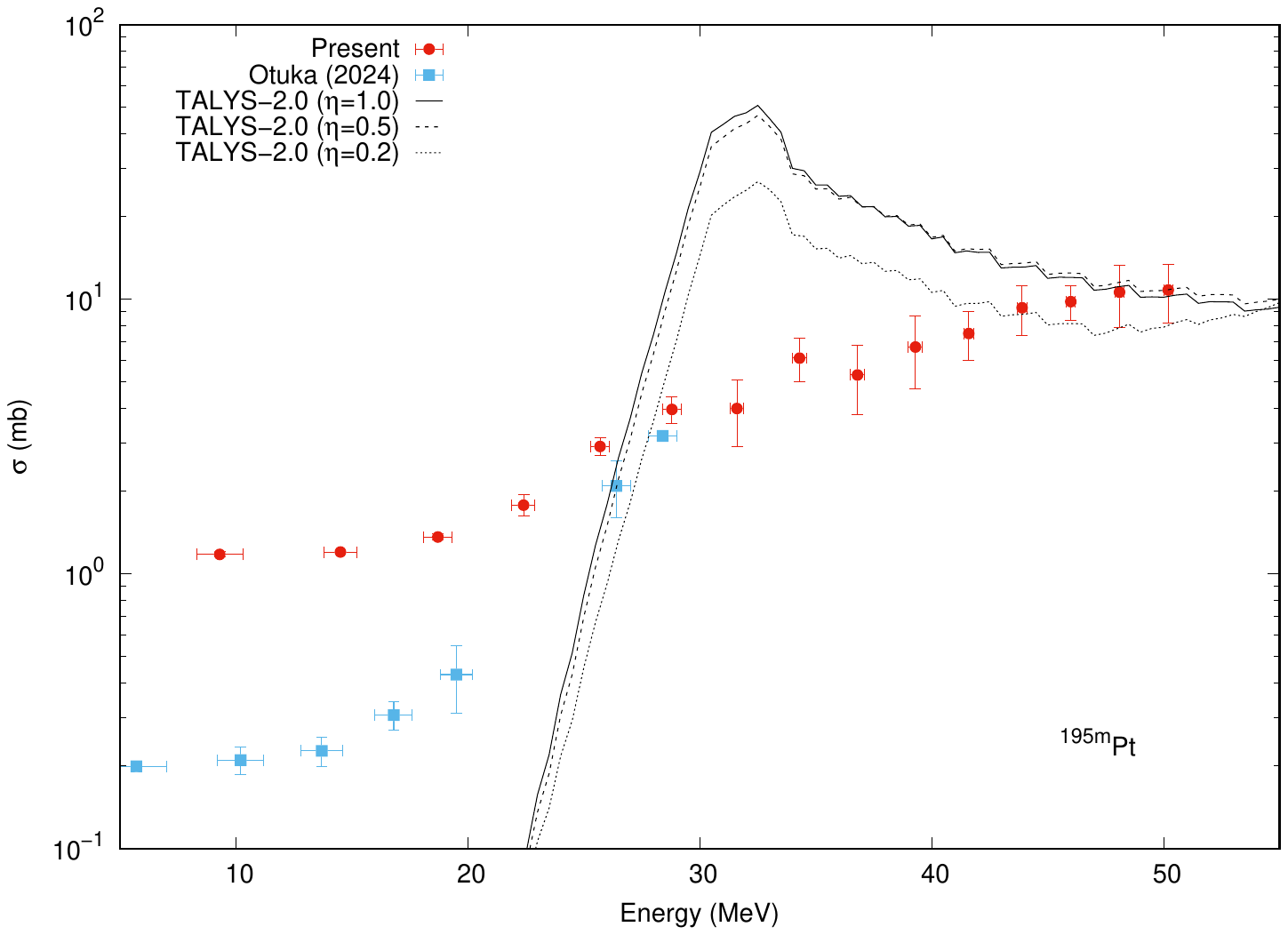}
\end{center}
\caption{
\natu Pt($\alpha$,x)$^{195m}$Pt cross sections.
Our present and previous experimental data points below $\sim$28~MeV should be seen with caution. See the main text for further details.
}
\label{fig:195mPt}
\end{figure}

In our previous study for irradiation of Pt foils by a 29~MeV alpha-particle beam~\cite{Otuka2024},
we observed the $^{195m}$Pt production cross section ($\sim$0.1 mb) in the low energy region below 20~MeV while TALYS-2.0 does not support it.
Figure~\ref{fig:195mPt} shows that this situation is more evident in our new data points from the 50~MeV alpha-particle irradiation,
and this could indicate the presence of $^{195}$Pt(n,n')$^{195m}$Pt reaction induced by secondary neutrons.
The cross section derived from decay curve analysis of the 98 keV gamma emission rate is shown as the \natu Pt($\alpha$,x)$^{195m}$Pt cross sections in this figure,
but it must be seen with caution.
We plan to perform a Monte Carlo simulation to estimate production of $^{195m}$Pt due to the $^{195}$Pt(n,n')$^{195m}$Pt reaction induced by the secondary neutrons.
%
%

\clearpage

\section{Summary}
We have determined the cross sections for production of various radionuclides from the interaction of Pt with an alpha-particle beam up to 50~MeV by using the stacked foil technique and off-line gamma spectroscopy.
We derived independent production cross sections for formation of 22 reaction products by performing decay curve analysis involving more than 40 gamma lines simultaneously.
The newly obtained cross sections agree with those measured by us previously up to 29~MeV and are also consistent with the literature data compiled in the EXFOR library except for $^{200m}$Au,$^{199}$Au and $^{197m,g}$Hg productions.
We demonstrated that the cross sections reported by Sagaidak \etal.~\cite{Sagaidak1979} become comparable with those reported by others if we interpreted the definition of their cross sections compiled in the EXFOR library properly.
Additionally, we confirmed their $^{198}$Au production cross section should not be treated as the $^{198g+m}$Au production cross section but as the $^{198g}$Au cumulative production cross section due to the close half-lives of the metastable and ground states.

The simultaneous decay curve analysis can utilize two gamma lines that are inseparable from each other without difficulty and provides independent production cross sections directly.
However, we had to exclude the gamma line associated with $^{193m}$Hg EC/$\beta^+$ decay and gamma lines interfering with them due to insufficient knowledge of the $^{193m}$Hg decay data.
We plan to investigate this situation further by analysing the spectra continuously measured for more than two months for the activity of a thick Pt foil irradiated by a 50~MeV alpha-particle beam.

The $^{195m}$Pt production cross section was measurable below $\sim$28~MeV contrary to prediction by TALYS-2.0,
and the secondary neutron effects such as $^{195}$Pt(n,n')$^{195m}$Pt reaction must be investigated to understand the situation.
In some cases (\eg, $^{198}$Au, $^{197}$Hg),
estimation of isomer production cross sections by TALYS-2.0 is improved by reducing the spin cutoff parameter from the one derived with the rigid body model.
But we found that it is not always the solution to improve description of the isomeric ratio (\eg, $^{196}$Au, $^{193}$Hg).

\begin{acknowledgement}
This work was carried out at RI Beam Factory operated by RIKEN Nishina Center and CNS, University of Tokyo, Japan.
The authors would like to thank Dr.~Y.~Shigekawa, Dr.~Y.~Kanayama, and Dr.~H.~Shimizu (RIKEN Nishina Center) for technical assistance with the experiment.
We would like to express our appreciation to Dr.~M.~Shamsuzzoha Basunia (Lawrence Berkeley National Laboratory) for his comments on the decay data of $^{193m}$Hg.
We also thank Dr.~Chushiro Yonezawa (The Japan Institute of International Affairs) for discussion on possible coincidence summing effects of two X-rays.
The IAEA LiveChart of Nuclides (https://nds.iaea.org/livechart/) developed and maintained by Dr.~Marco Verpelli (IAEA Nuclear Data Section) has been an essential tool for extraction of the decay data from the ENSDF library.
This research was partially performed by the commissioned research funds provided by F-REI (JPFR23040201 and 24040201).
\end{acknowledgement}

\clearpage

\appendix*
\section{Excitation functions not measured by others}
\begin{figure}[hbtp]
\begin{center}
\includegraphics[width=0.8\textwidth]{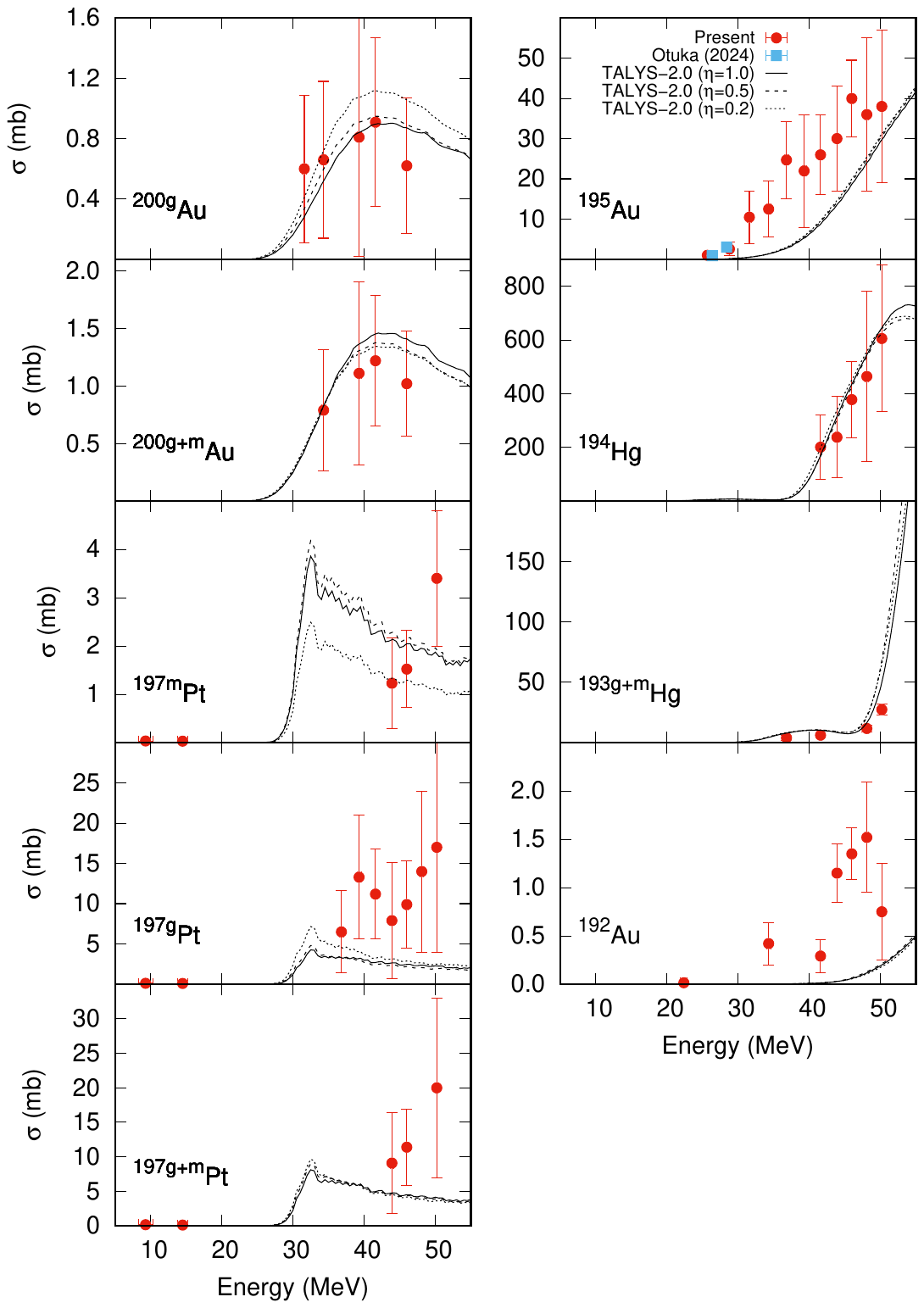}
\end{center}
\caption{
\natu Pt($\alpha$,x)$^{200g,g+m}$Au, $^{197m,g,g+m}$Pt, $^{195}$Au, $^{194}$Hg, $^{193g+m}$Hg and $^{192}$Au cross sections.
}
\label{fig:appendix-xs}
\end{figure}

\begin{figure}[hbtp]
\begin{center}
\includegraphics[width=0.8\textwidth]{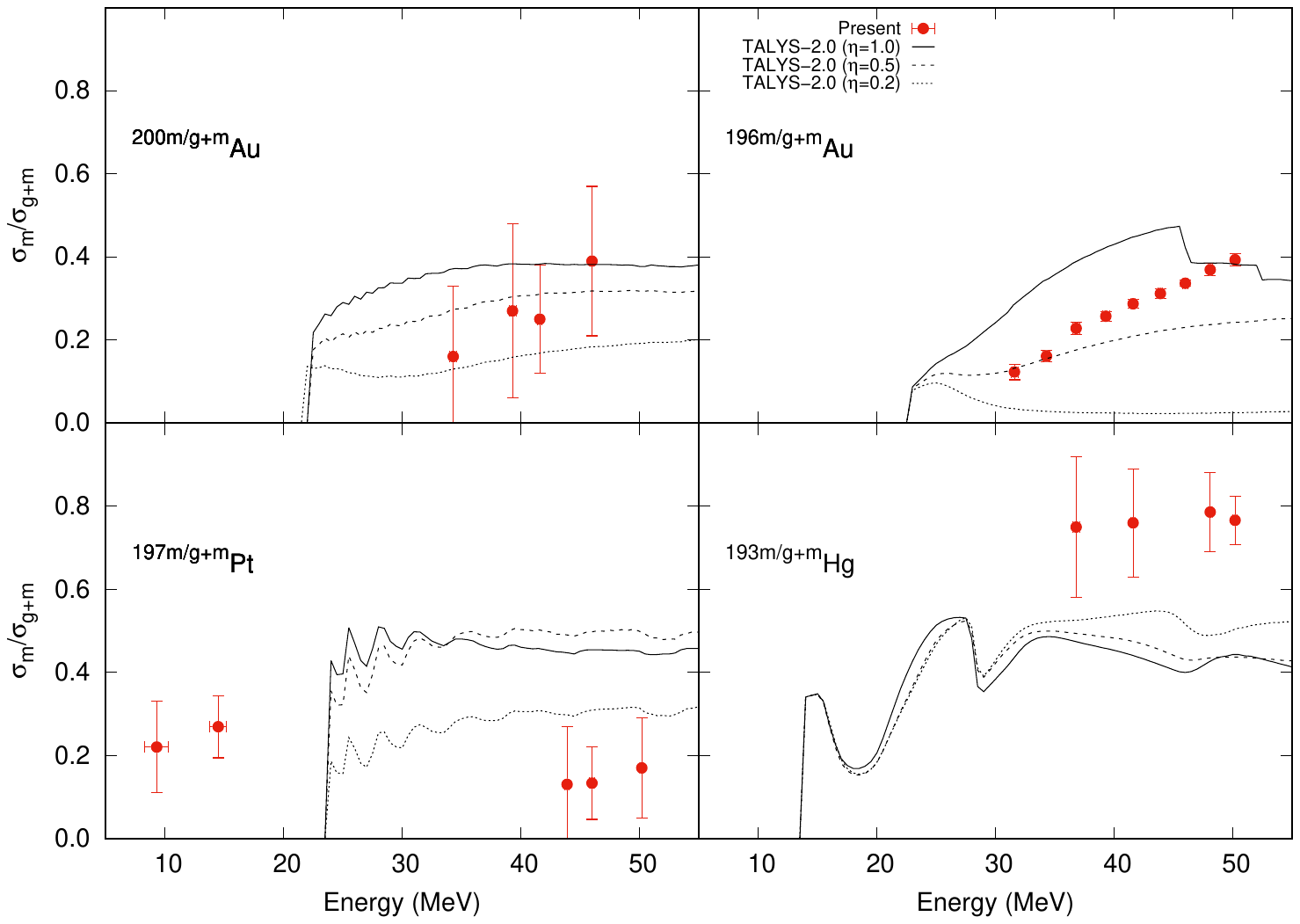}
\end{center}
\caption{
\natu Pt($\alpha$,x)$^{200}$Au, $^{197}$Pt, $^{196}$Au and $^{193}$Hg isomeric ratios.
}
\label{fig:appendix-ir}
\end{figure}

Figures~\ref{fig:appendix-xs} and \ref{fig:appendix-ir} show the production cross sections and isomeric ratios determined by us in the present and previous works but not measured by other groups.
Production of $^{197}$Pt seen in the low energy region may include contribution from the $^{198}$Pt(n,2n)$^{197}$Pt reaction.

\section{Table of newly measured excitation functions}
Table~\ref{tab:appendix} collects the numerical data of the production cross sections and isomeric ratios determined in the present work.
The total ($g+m$) cross sections and isomeric ratios (IR) were not measured directly,
but derived from the partial ($g$ and $m$) cross sections,
which are the measurables in the present work.
All cross sections in the tables are for natural Pt samples and not for isotopic Pt samples.

\begin{longtable}{cccccccccc}
\caption{
Production cross section (mb) and isomeric ratio $IR$ = $\sigma_m/(\sigma_g+\sigma_m)$ at the incident energy $E$ (MeV).
$\Delta \sigma$ gives the uncertainty in the cross section propagated from the uncertainty in the peak areas.
An asterisk indicates presence of a value with uncertainty higher than 50\%.
The $^{195m}$Pt production cross sections below $\sim$28~MeV are italicised as they should be seen with caution as discussed in Sec.~\ref{sec:195mPt}.
}
\label{tab:appendix}
\\
\hline
\hline
 &  & \multicolumn{2}{c}{$^{200m}$Au }  & \multicolumn{2}{c}{$^{200g}$Au }  & \multicolumn{2}{c}{$^{200g+m}$Au }  & \multicolumn{2}{c}{$^{200}$Au IR }  \\
\hline
$E$ & $\Delta E$ & $\sigma$ & $\Delta \sigma$ & $\sigma$ & $\Delta \sigma$ & $\sigma$ & $\Delta \sigma$ & $IR$ & $\Delta IR$\\
\hline
50.2 & 0.2 & 0.56 & 0.23 &  &  &  &  &  &  \\
48.1 & 0.2 & 0.48 & 0.24 &  &  &  &  &  &  \\
46.0 & 0.2 & 0.403 & 0.099 & * & * & 1.02 & 0.46 & 0.39 & 0.18 \\
43.9 & 0.2 & 0.43 & 0.15 &  &  &  &  &  &  \\
41.6 & 0.2 & 0.31 & 0.11 & * & * & 1.22 & 0.57 & * & * \\
39.3 & 0.3 & 0.30 & 0.13 & * & * & * & * & * & * \\
36.8 & 0.3 & 0.37 & 0.11 &  &  &  &  &  &  \\
34.3 & 0.3 & * & * & * & * & * & * & * & * \\
31.6 & 0.3 &  &  & * & * &  &  &  &  \\
\hline
\hline
 &  & \multicolumn{2}{c}{$^{198m}$Au }  & \multicolumn{2}{c}{$^{198g}$Au }  & \multicolumn{2}{c}{$^{198g+m}$Au }  & \multicolumn{2}{c}{$^{198}$Au IR }  \\
\hline
$E$ & $\Delta E$ & $\sigma$ & $\Delta \sigma$ & $\sigma$ & $\Delta \sigma$ & $\sigma$ & $\Delta \sigma$ & $IR$ & $\Delta IR$\\
\hline
50.2 & 0.2 & 3.09 & 0.20 & 4.80 & 0.52 & 7.89 & 0.56 & 0.392 & 0.030 \\
48.1 & 0.2 & 2.72 & 0.21 & 4.72 & 0.53 & 7.44 & 0.57 & 0.366 & 0.032 \\
46.0 & 0.2 & 2.14 & 0.10 & 6.04 & 0.28 & 8.18 & 0.30 & 0.262 & 0.013 \\
43.9 & 0.2 & 2.06 & 0.13 & 5.28 & 0.37 & 7.34 & 0.39 & 0.281 & 0.019 \\
41.6 & 0.2 & 1.80 & 0.10 & 5.51 & 0.28 & 7.31 & 0.30 & 0.246 & 0.014 \\
39.3 & 0.3 & 1.64 & 0.14 & 5.63 & 0.43 & 7.27 & 0.45 & 0.226 & 0.020 \\
36.8 & 0.3 & 1.260 & 0.098 & 5.36 & 0.27 & 6.62 & 0.29 & 0.190 & 0.014 \\
34.3 & 0.3 & 0.986 & 0.073 & 4.06 & 0.19 & 5.05 & 0.20 & 0.195 & 0.014 \\
31.6 & 0.3 & 0.445 & 0.068 & 2.90 & 0.18 & 3.35 & 0.19 & 0.133 & 0.019 \\
28.8 & 0.4 & 0.163 & 0.043 & 1.59 & 0.11 & 1.75 & 0.12 & 0.093 & 0.023 \\
25.7 & 0.4 &  &  & 0.634 & 0.046 &  &  &  &  \\
22.4 & 0.5 &  &  & 0.070 & 0.010 &  &  &  &  \\
9.3 & 1.0 &  &  & 0.00256 & 0.00091 &  &  &  &  \\
\hline
\hline
 &  & \multicolumn{2}{c}{$^{197m}$Hg }  & \multicolumn{2}{c}{$^{197g}$Hg }  & \multicolumn{2}{c}{$^{197g+m}$Hg }  & \multicolumn{2}{c}{$^{197}$Hg IR }  \\
\hline
$E$ & $\Delta E$ & $\sigma$ & $\Delta \sigma$ & $\sigma$ & $\Delta \sigma$ & $\sigma$ & $\Delta \sigma$ & $IR$ & $\Delta IR$\\
\hline
50.2 & 0.2 & 102.2 & 7.7 & * & * & 148 & 35 & 0.69 & 0.16 \\
48.1 & 0.2 & 78.5 & 4.4 & * & * & 124 & 33 & 0.64 & 0.17 \\
46.0 & 0.2 & 88.6 & 2.1 & * & * & 117 & 16 & 0.76 & 0.10 \\
43.9 & 0.2 & 98.2 & 8.1 & * & * & 127 & 26 & 0.77 & 0.15 \\
41.6 & 0.2 & 152.4 & 4.6 & * & * & 176 & 20 & 0.864 & 0.093 \\
39.3 & 0.3 & 217.0 & 8.4 & 70 & 29 & 287 & 30 & 0.756 & 0.077 \\
36.8 & 0.3 & 281.2 & 7.2 & 105 & 25 & 386 & 26 & 0.728 & 0.047 \\
34.3 & 0.3 & 286.2 & 5.7 & 118 & 17 & 404 & 18 & 0.708 & 0.030 \\
31.6 & 0.3 & 245.8 & 5.5 & 123 & 18 & 369 & 19 & 0.666 & 0.033 \\
28.8 & 0.4 & 223.7 & 4.3 & 96 & 12 & 320 & 13 & 0.700 & 0.027 \\
25.7 & 0.4 & 157.5 & 3.1 & 110 & 9.4 & 267.5 & 9.9 & 0.589 & 0.021 \\
22.4 & 0.5 & 65.23 & 0.86 & 73.6 & 3.3 & 138.8 & 3.4 & 0.470 & 0.012 \\
18.7 & 0.6 & 5.25 & 0.11 & 11.22 & 0.54 & 16.47 & 0.55 & 0.319 & 0.011 \\
14.5 & 0.7 &  &  & * & * &  &  &  &  \\
\hline
\hline
 &  & \multicolumn{2}{c}{$^{197m}$Pt }  & \multicolumn{2}{c}{$^{197g}$Pt }  & \multicolumn{2}{c}{$^{197g+m}$Pt }  & \multicolumn{2}{c}{$^{197}$Pt IR }  \\
\hline
$E$ & $\Delta E$ & $\sigma$ & $\Delta \sigma$ & $\sigma$ & $\Delta \sigma$ & $\sigma$ & $\Delta \sigma$ & $IR$ & $\Delta IR$\\
\hline
50.2 & 0.2 & 3.4 & 1.4 & * & * & * & * & * & * \\
48.1 & 0.2 &  &  & * & * &  &  &  &  \\
46.0 & 0.2 & * & * & * & * & 11.4 & 5.5 & * & * \\
43.9 & 0.2 & * & * & * & * & * & * & * & * \\
41.6 & 0.2 &  &  & * & * &  &  &  &  \\
39.3 & 0.3 &  &  & * & * &  &  &  &  \\
36.8 & 0.3 &  &  & * & * &  &  &  &  \\
14.5 & 0.7 & 0.0291 & 0.0068 & 0.079 & 0.024 & 0.108 & 0.025 & 0.269 & 0.075 \\
9.3 & 1.0 & * & * & 0.111 & 0.029 & 0.143 & 0.035 & * & * \\
\hline
\hline
 &  & \multicolumn{2}{c}{$^{196m}$Au }  & \multicolumn{2}{c}{$^{196g}$Au }  & \multicolumn{2}{c}{$^{196g+m}$Au }  & \multicolumn{2}{c}{$^{196}$Au IR }  \\
\hline
$E$ & $\Delta E$ & $\sigma$ & $\Delta \sigma$ & $\sigma$ & $\Delta \sigma$ & $\sigma$ & $\Delta \sigma$ & $IR$ & $\Delta IR$\\
\hline
50.2 & 0.2 & 17.67 & 0.82 & 27.3 & 1.0 & 45.0 & 1.3 & 0.393 & 0.014 \\
48.1 & 0.2 & 14.91 & 0.77 & 25.53 & 0.97 & 40.4 & 1.2 & 0.369 & 0.015 \\
46.0 & 0.2 & 12.57 & 0.35 & 24.79 & 0.45 & 37.36 & 0.57 & 0.3365 & 0.0074 \\
43.9 & 0.2 & 10.08 & 0.45 & 22.27 & 0.57 & 32.35 & 0.73 & 0.312 & 0.011 \\
41.6 & 0.2 & 7.70 & 0.33 & 19.10 & 0.42 & 26.80 & 0.53 & 0.2873 & 0.0099 \\
39.3 & 0.3 & 5.50 & 0.30 & 15.87 & 0.43 & 21.37 & 0.52 & 0.257 & 0.012 \\
36.8 & 0.3 & 3.66 & 0.30 & 12.41 & 0.37 & 16.07 & 0.48 & 0.228 & 0.015 \\
34.3 & 0.3 & 1.74 & 0.17 & 9.09 & 0.23 & 10.83 & 0.29 & 0.161 & 0.014 \\
31.6 & 0.3 & 0.77 & 0.13 & 5.49 & 0.19 & 6.26 & 0.23 & 0.123 & 0.019 \\
28.8 & 0.4 &  &  & 2.794 & 0.075 &  &  &  &  \\
25.7 & 0.4 &  &  & 0.624 & 0.034 &  &  &  &  \\
22.4 & 0.5 & * & * &  &  &  &  &  &  \\
18.7 & 0.6 &  &  & 0.0109 & 0.0053 &  &  &  &  \\
14.5 & 0.7 &  &  & 0.00955 & 0.00089 &  &  &  &  \\
\hline
\hline
 &  & \multicolumn{2}{c}{$^{195m}$Hg }  & \multicolumn{2}{c}{$^{195g}$Hg }  & \multicolumn{2}{c}{$^{195g+m}$Hg }  & \multicolumn{2}{c}{$^{195}$Hg IR }  \\
\hline
$E$ & $\Delta E$ & $\sigma$ & $\Delta \sigma$ & $\sigma$ & $\Delta \sigma$ & $\sigma$ & $\Delta \sigma$ & $IR$ & $\Delta IR$\\
\hline
50.2 & 0.2 & 518.2 & 8.4 & 64.7 & 6.2 & 583 & 10 & 0.8890 & 0.0096 \\
48.1 & 0.2 & 520.1 & 8.8 & 60.8 & 6.0 & 581 & 11 & 0.8953 & 0.0094 \\
46.0 & 0.2 & 529.6 & 4.4 & 66.6 & 3.1 & 596.2 & 5.4 & 0.8883 & 0.0047 \\
43.9 & 0.2 & 540.9 & 6.1 & 70.2 & 4.3 & 611.1 & 7.5 & 0.8851 & 0.0063 \\
41.6 & 0.2 & 537.4 & 4.8 & 75.7 & 3.4 & 613.1 & 5.9 & 0.8765 & 0.0050 \\
39.3 & 0.3 & 489.1 & 6.4 & 83.4 & 4.8 & 572.5 & 8.0 & 0.8543 & 0.0073 \\
36.8 & 0.3 & 374.9 & 4.5 & 75.5 & 3.0 & 450.4 & 5.4 & 0.8324 & 0.0058 \\
34.3 & 0.3 & 276.7 & 3.2 & 65.6 & 2.6 & 342.3 & 4.1 & 0.8084 & 0.0064 \\
31.6 & 0.3 & 155.9 & 2.8 & 49.1 & 2.0 & 205.0 & 3.4 & 0.7605 & 0.0081 \\
28.8 & 0.4 & 29.7 & 1.2 & 12.96 & 0.86 & 42.7 & 1.5 & 0.696 & 0.016 \\
25.7 & 0.4 &  &  &  &  &  &  &  &  \\
22.4 & 0.5 & 0.440 & 0.079 & 0.393 & 0.076 & 0.83 & 0.11 & 0.528 & 0.066 \\
18.7 & 0.6 & 0.093 & 0.031 & 0.113 & 0.024 & 0.206 & 0.039 & 0.451 & 0.098 \\
\hline
\hline
 &  & \multicolumn{2}{c}{$^{193m}$Hg }  & \multicolumn{2}{c}{$^{193g}$Hg }  & \multicolumn{2}{c}{$^{193g+m}$Hg }  & \multicolumn{2}{c}{$^{193}$Hg IR }  \\
\hline
$E$ & $\Delta E$ & $\sigma$ & $\Delta \sigma$ & $\sigma$ & $\Delta \sigma$ & $\sigma$ & $\Delta \sigma$ & $IR$ & $\Delta IR$\\
\hline
50.2 & 0.2 & 20.9 & 4.3 & 6.4 & 1.6 & 27.3 & 4.6 & 0.766 & 0.058 \\
48.1 & 0.2 & 9.2 & 2.8 & 2.5 & 1.2 & 11.7 & 3.0 & 0.786 & 0.095 \\
46.0 & 0.2 &  &  & * & * &  &  &  &  \\
43.9 & 0.2 &  &  & * & * &  &  &  &  \\
41.6 & 0.2 & 4.7 & 2.0 & * & * & 6.2 & 2.2 & 0.76 & 0.13 \\
39.3 & 0.3 & * & * &  &  &  &  &  &  \\
36.8 & 0.3 & * & * & * & * & * & * & 0.75 & 0.17 \\
34.3 & 0.3 & 4.1 & 1.4 &  &  &  &  &  &  \\
31.6 & 0.3 &  &  & * & * &  &  &  &  \\
\hline
\hline
 &  & \multicolumn{2}{c}{$^{199m}$Hg }  & \multicolumn{2}{c}{$^{199}$Au }  & \multicolumn{2}{c}{$^{195}$Au }  & \multicolumn{2}{c}{$^{195m}$Pt }  \\
\hline
$E$ & $\Delta E$ & $\sigma$ & $\Delta \sigma$ & $\sigma$ & $\Delta \sigma$ & $\sigma$ & $\Delta \sigma$ & $\sigma$ & $\Delta \sigma$\\
\hline
50.2 & 0.2 & 7.97 & 0.59 & 5.99 & 0.42 & 38 & 19       & 10.8          & 2.6           \\
48.1 & 0.2 & 9.37 & 0.70 & 7.98 & 0.52 & * & *         & 10.6          & 2.7           \\
46.0 & 0.2 & 11.70 & 0.39 & 7.91 & 0.25 & 40.0 & 9.5   &  9.8          & 1.4           \\
43.9 & 0.2 & 15.08 & 0.62 & 8.01 & 0.36 & 30 & 13      &  9.3          & 1.9           \\
41.6 & 0.2 & 21.15 & 0.63 & 7.43 & 0.28 & 26 & 10      &  7.5          & 1.5           \\
39.3 & 0.3 & 29.27 & 0.99 & 6.55 & 0.37 & * & *        &  6.7          & 2.0           \\
36.8 & 0.3 & 58.2 & 1.2 & 5.70 & 0.27 & 24.7 & 9.6     &  5.3          & 1.5           \\
34.3 & 0.3 & 73.3 & 1.6 & 4.36 & 0.18 & * & *          &  6.1          & 1.1           \\
31.6 & 0.3 & 67.0 & 1.5 & 3.05 & 0.19 & * & *          &  4.0          & 1.1           \\
28.8 & 0.4 & 43.6 & 1.1 & 1.96 & 0.12 & * & *          & \textit{3.97} & \textit{0.45} \\
25.7 & 0.4 & 8.26 & 0.42 & 0.857 & 0.071 & 1.01 & 0.17 & \textit{2.91} & \textit{0.21} \\
22.4 & 0.5 & 1.531 & 0.091 & 0.090 & 0.025 &  &        & \textit{1.78} & \textit{0.16} \\
18.7 & 0.6 & 0.859 & 0.045 & 0.0125 & 0.0042 &  &      & \textit{1.363}& \textit{0.039} \\
14.5 & 0.7 &  &  & 0.0174 & 0.0014 &  &                & \textit{1.201}& \textit{0.012} \\
9.3 & 1.0 &  &  & 0.0181 & 0.0026 &  &                 & \textit{1.178}& \textit{0.028} \\
2.2 & 0.8 & 14.7 & 4.7 &  &  &  &                      & \textit{1.152}& \textit{0.052} \\
\hline
\hline
 &  & \multicolumn{2}{c}{$^{194}$Hg }  & \multicolumn{2}{c}{$^{194}$Au }  & \multicolumn{2}{c}{$^{192}$Hg }  & \multicolumn{2}{c}{$^{192}$Au }  \\
\hline
$E$ & $\Delta E$ & $\sigma$ & $\Delta \sigma$ & $\sigma$ & $\Delta \sigma$ & $\sigma$ & $\Delta \sigma$ & $\sigma$ & $\Delta \sigma$\\
\hline
50.2 & 0.2 & 606 & 273 & 14.81 & 0.70 & 10.87 & 0.53 & * & * \\
48.1 & 0.2 & * & * & 9.46 & 0.43 & 7.94 & 0.53 & 1.52 & 0.57 \\
46.0 & 0.2 & 378 & 142 & 5.73 & 0.26 & 6.23 & 0.25 & 1.35 & 0.27 \\
43.9 & 0.2 & * & * & 3.06 & 0.30 & 2.60 & 0.23 & 1.15 & 0.30 \\
41.6 & 0.2 & * & * & 1.82 & 0.21 & 1.26 & 0.11 & * & * \\
39.3 & 0.3 &  &  & 1.09 & 0.25 &  &  &  &  \\
36.8 & 0.3 &  &  & 0.63 & 0.17 &  &  &  &  \\
34.3 & 0.3 &  &  &  &  &  &  & * & * \\
31.6 & 0.3 &  &  &  &  & * & * &  &  \\
28.8 & 0.4 &  &  &  &  & 0.124 & 0.030 &  &  \\
25.7 & 0.4 &  &  &  &  & 0.081 & 0.021 &  &  \\
22.4 & 0.5 &  &  & 0.0369 & 0.0097 & * & * & * & * \\
18.7 & 0.6 &  &  & 0.0346 & 0.0031 &  &  &  &  \\
\hline
\hline
\end{longtable}

\clearpage

\bibliography{Pt_a}
\end{document}